\documentclass[a4paper,11pt]{article}
\usepackage{jcappub} 
\usepackage{soul}
\usepackage{graphicx}
\usepackage{caption}
\usepackage{subcaption}
\usepackage{mathrsfs}

\usepackage{siunitx}
\let\svqty\qty
\usepackage{physics}
\let\qty\svqty
\usepackage{amsmath,amssymb,amsfonts}
\usepackage{tensor}

\def\be#1\ee{\begin{align}#1\end{align}} 
\def\bse#1\ese{\begin{subequations}#1\end{subequations}} 
\newcommand{\iu}{\text{i}} 
\newcommand{\Lie}[2]{\ensuremath{\pounds_{#1} #2 }} 
\newcommand{\htheta}{\ensuremath{ { \hat{\theta}} } } 
\newcommand{\hphi}{\ensuremath{ { \hat{\varphi}} } } 

\newcommand{\quh}{ {\textsc{quh} } } 
\newcommand{\rmd}{\mathrm{d}} 
\newcommand{\qu}[1]{``{#1}"} 

\usepackage{hyperref}
\usepackage[capitalize]{cleveref}
\crefname{equation}{eq.}{eqs.}
\crefname{figure}{figure}{figures}
\crefname{table}{table}{tables}
\crefname{subequation}{eqs.}{eqs.}
\labelcrefformat{subequation}{#2(#1)#3}
\crefformat{subequations}{#2(#1)#3}
\crefname{section}{section}{sections}
\crefname{appendix}{appendix}{appendices}

\usepackage{soul}
\usepackage{comment}

\author[a]{Francesco Del Porro}
\affiliation[a]{Niels Bohr International Academy, Niels Bohr Institute, Blegdamsvej 17, 2100 Copenhagen, Denmark}
\emailAdd{francesco.del.porro@nbi.ku.dk}

\author[b,c,d]{, Stefano Liberati}
\affiliation[b]{SISSA, International School for Advanced Studies, via Bonomea 265, 34136 Trieste, Italy}
\affiliation[c]{INFN, Sezione di Trieste, via Valerio 2, 34127 Trieste, Italy}
\affiliation[d]{IFPU, Institute for Fundamental Physics of the Universe, via Beirut 2, 34014 Trieste, Italy}
\emailAdd{liberati@sissa.it}

\author[e]{, and Jacopo Mazza}
\affiliation[e]{Université Paris-Saclay, CNRS/IN2P3, IJCLab, 91405 Orsay, France}
\emailAdd{jacopo.mazza@ijclab.in2p3.fr}

\title{\boldmath Universal Horizons without Hypersurface Orthogonality}

\abstract{
A key consequence of Lorentz-violating gravity is the emergence of modified dispersion relations implying the absence of a universal maximum propagation speed. 
This challenges the conventional notion of the event horizon as a causal boundary common to all degrees of freedom. 
However, certain solutions in these theories exhibit {\em universal horizons} --- surfaces capable of trapping signals of arbitrarily high speed, thereby restoring the notion of black hole. 
Previous studies have extensively characterised universal horizons in settings where Lorentz violation is encoded via a hypersurface-orthogonal \ae{}ther. 
In this work, we explore the possibility of extending this concept to more general cases where hypersurface orthogonality is relaxed. 
To do so, we construct a candidate trapping surface and analyse its causal properties using a general model for Lorentz-violating matter. 
We find that, in addition to the standard conditions associated to universal horizons, a local vanishing of the \ae{}ther's twist is also necessary. 
We then provide an explicit example of such a universal horizon by suitably deforming the \ae{}ther flow in a stealth Kerr solution recently found in Einstein--\ae{}ther theory.
Moreover, we analyse the behaviour of trajectories which are not analytical at the universal horizon and discuss the implications of our findings for Hawking radiation. 
While our analysis is motivated by Einstein-\ae{t}her gravity, our results apply to broader classes of Lorentz-violating theories, further supporting the relevance of black hole phenomenology in these frameworks.
}

\begin{document}
\maketitle
\flushbottom

\section{Introduction}\label{s:intro}

Lorentz invariance is a cornerstone of modern physics and, as such, it has been extensively tested.
This fact notwithstanding, reasons to envisage violations of Lorentz invariance abound:
on the one hand, several approaches to quantum gravity point to the possibility that Lorentz invariance might be violated at a fundamental level or deformed at very high energies \cite{MattinglyModernTests2005,LiberatiTestsLorentz2013,Amelino-CameliaQuantumSpacetime2013,AddaziQuantumGravity2022};
on the other, the spontaneous breaking of Lorentz invariance is fairly common in solutions to modified theories of gravity, even at low energies \cite{Arkani-HamedGhostCondensation2004,MukohyamaEffectiveField2022,MukohyamaQuasinormalModes2023,CardosoBlackHoles2024,LangloisBlackHole2021,LangloisAsymptoticsLinear2021,LangloisEffectiveMetric2022}.

A hallmark of the violation of Lorentz invariance is the appearance of modifications in the dispersion relations of propagating degrees of freedom.
The simplest form of such modifications is a shift in the propagation speed of massless modes in vacuum, since lacking Lorentz invariance these modes are no longer bound to the light cone specified by the spacetime metric.
A less trivial possibility is that the dispersion relations acquire a nonlinear character, which entails that, formally, propagation speeds have no upper bound.
In either case, causal signals propagating faster than light may be allowed.

Evidently, violations of Lorentz invariance are exceptionally consequential for black holes, since horizons as specified by the spacetime metric might lose their interpretation as universal causal boundaries.
In situations in which a finite maximum propagation speed persists, one might still be able to identify regions of no escape to rightfully call black holes --- although the ensuing phenomenology appears far from trivial \cite{DubovskySpontaneousBreaking2006,ElingLorentzViolation2007,JacobsonBlackHole2010,BenkelDynamicalObstruction2018,CardosoBlackHoles2024}. 
However, if degrees of freedom with nonlinear superluminal dispersion relations exist, the very notion of black hole becomes \emph{seemingly} void, since infinite-speed signals would be able to causally connect any region of spacetime to infinity.

Surprisingly, a careful analysis of specific solutions to some theories of gravity with broken Lorentz invariance revealed the existence of a peculiar surface that can trap degrees of freedom of arbitrarily high speed.
Such surface therefore constitutes a causal boundary that is  universal to all dispersion relations, and for this reason it has been dubbed \emph{universal horizon} \cite{eling_black_2006,BlasHoravaGravity2011,BarausseBlackHoles2011,BarausseBlackHoles2013,SaravaniDynamicalEmergence2014,BhattacharyyaUniversalHorizons2014,barausse_slowly_2016,FranchiniBlackHole2017}.

Universal horizons are distinct and largely independent from metric horizons, and yet the two display remarkable similarities: for example, universal horizons have been shown to satisfy some laws of black hole mechanics \cite{BerglundMechanicsUniversal2012,Pacilio:2017emh,Pacilio:2017swi,MartinMixed2024} and to exhibit thermal properties \cite{del_porro_time_2022,del_porro_gravitational_2022,del_porro_hawking_2023}.
They emerge naturally in the context of globally foliated manifolds, whereby the breaking of Lorentz invariance is realised through the appearance of a preferred slicing of the spacetime in terms of hypersurfaces of simultaneity \cite{LinUniversalHorizons2014,bhattacharyya_causality_2016}.

Indeed, the first examples of universal horizons have been identified in khronometric theory \cite{BlasModelsNonrelativistic2011}, a covariant scalar--tensor theory of gravity that coincides with the low-energy limit of nonprojectable Ho\v{r}ava gravity \cite{HoravaQuantumGravity2009,BlasExtraMode2009,BlasModelsNonrelativistic2011,Herrero-ValeaStatusHorava2023};
but similar instances could be found, for instance, in higher-order scalar--tensor theories in which the scalar has an everywhere-timelike gradient \cite{Arkani-HamedGhostCondensation2004,MukohyamaEffectiveField2022,MukohyamaQuasinormalModes2023}.\footnote{Khronometric theory itself belongs to the U-DHOST class of higher-order scalar-tensor theories of gravity \cite{DeFeliceShadowyModes2018,DeFeliceNonlinearDefinition2021,DeFeliceAvoidanceStrong2022}.}

Any global foliation admits a dual description in terms of the vector field that is everywhere orthogonal to it.
Conversely, by Frobenius' theorem, a vector field is orthogonal to a family of codimension-one hypersurfaces if, and only if, its twist tensor vanishes identically everywhere in the manifold. 
In the context of Lorentz-violating theories of gravity, the foliation is typically taken to be spacelike and the vector field is then interpreted as providing a preferred timelike threading, i.e.~a preferred timelike direction in the tangent space.

However, in many cases the assumption of hypersurface orthogonality is dropped and one considers theories with only a preferred threading, not necessarily related to a preferred foliation. 
This is the case of the so-called Einstein--\ae{}ther gravity \cite{jacobson_gravity_2001,jacobson_einstein-aether_2008,JacobsonExtendedHorava2010,jacobson_undoing_2014}: a vector--tensor extension of general relativity characterised by a constraint ensuring that the vector is everywhere timelike and of unit norm, although not necessarily hypersurface orthogonal, so that the vector's twist is dynamical and generically nonzero. 
Einstein--\ae{}ther is closely related to the aforementioned khronometric theory: indeed  (with some caveats \cite{barausse_slowly_2016}) the latter can be regarded as a \qu{hypersurface-orthogonal version} of the former \cite{jacobson_undoing_2014}.

These remarks raise an important question:
Can the notion of universal horizon be extended to cases in which hypersurface orthogonality is absent? 
The goal of this article is to address this very question.
We shall establish under what conditions the answer is in the positive, and describe situations in which such conditions are only partially met. 
Moreover, we shall characterise the semiclassical radiative properties of universal horizons, highlighting the actual role played by hypersurface orthogonality.
In doing so, we will draw heavily on some previous characterisations of universal horizons, particularly that of \cite{bhattacharyya_causality_2016} and \cite{carballo-rubio_geodesically_2022}, developed for the hypersurface-orthogonal case.

We will also make frequent reference to the specific example of Einstein--\ae{}ther theory, and particularly to the rotating stealth Kerr solution of \cite{franzin_kerr_2023}, although our conclusions apply to much more general cases.

\bigskip

The signature of the spacetime metric is $(-,+,+,+)$.
We will assume we work in a local coordinate chart whenever necessary and generically use Latin letters to denote spacetime indices;
we will use Greek indices, instead, to refer to the stealth Kerr example of \cite{franzin_kerr_2023}, written in Boyer--Lindquist coordinates.
Contractions between vectors will often be denoted through the symbol \qu{$\cdot$}.

\section{Geometrical setup}\label{s:defs}

We start our discussion by laying out the geometrical setup we employ in the rest of the paper.
First we introduce the preferred frame responsible for the breaking of local Lorentz invariance, specifying the symmetries we assume it obeys.
Then we describe the new notion of causality induced by such preferred frame.
Finally, we introduce a set of constants of the motion, whose conservation descends from the symmetries.

\subsection{Preferred frame and related quantities}\label{s:pframe}

Our formalism is largely inspired by that of \cite{bhattacharyya_causality_2016}.
We consider covariant theories of gravity written in terms of a Lorentzian metric $g_{ab}$ (and possibly other fields), and assume that Lorentz invariance is broken by a (co)vector field $u_a$.
Such vector field, which we will call \emph{\ae{}ther}, is assumed to be everywhere timelike and normalised:
\be\label{eq:unorm}
g^{ab} u_a u_b = - 1\, .
\ee
Although this is not strictly necessary for the ensuing discussion, the situations we have in mind are such that the \ae{}ther is one of the dynamical fields in the theory, in the sense that it satisfies suitable evolution equations. 
In other words, we implicitly assume that the \ae{}ther is not a background structure, and the theories we consider are background independent.

We wish to focus on situations displaying a high degree of symmetry, akin to that of the Kerr spacetime, namely stationarity and axisymmetry.
We thus assume that the metric admits two Killing vector fields: $\chi^a$, generating time translations, and $\psi^a$, generating rotations around an axis; also, $\chi^a$ is assumed to be timelike in a neighbourhood of infinity, while $\psi^a$, whose orbits are closed, is assumed to be everywhere spacelike.

Moreover, we assume that the Killing vectors commute and that the \ae{}ther obeys the same symmetries as the metric, i.e.~we assume it to be Lie-dragged along the Killing vector fields:
\be 
&\Lie{\chi}{\psi^a} = 0\, ,  \label{eq:chiLiepsi}\\
\Lie{\chi}{u_a} =& 0 \qq{and} \Lie{\psi}{u_a} = 0 \, . \label{eq:uLie}
\ee
Clearly, one could consider different situations in which, for example, the \ae{}ther does not obey the same symmetries as the metric.
However, the assumptions of \cref{eq:uLie} are consistent with the idea that the metric--\ae{}ther configuration under consideration be a stationary and axisymmetric solution to a suitable set of equations of motion.

In addition, we assume that 
\be\label{eq:uZAMO}
\left( u \cdot \psi \right) = 0\, .
\ee
This condition, which entails that an observer at rest in the \ae{}ther frame has zero Killing angular momentum (i.e.~it is \qu{ZAMO}), is herein assumed because it greatly simplifies the discussion.
Though this does not seem indispensable, it is physically well motivated by the requirement that there exist no closed causal curves --- where the precise meaning of \qu{causal} will become clear momentarily.
Similarly, our arguments do not require the assumption that the spacetime be asymptotically flat and that 
\be\label{eq:asyflat}
\left(u \cdot \chi \right) \to - 1 \qq{at} \mathfrak{i}^0 \,;
\ee
however, this is the common practice to which we will implicitly conform.

It will often be convenient to express tensors in terms of a basis that is adapted to the frame provided by $u_a$.
We thus complement $u_a$ with a triad $\{s_a,\ \htheta_a,\ \hphi_a \}$ of mutually orthogonal, unit-norm, spacelike vectors that we choose as follows:
we take $\hphi_a$ in the direction of the Killing vector $\psi_a$, i.e.
\be
\hphi_a := \frac{\psi_a}{\sqrt{\left( \psi \cdot \psi \right)}}\, ;
\ee
we then take $\htheta_a$ to be orthogonal to the other Killing vector $\chi_a$, so that $( \htheta \cdot \chi ) =( \htheta \cdot u ) = ( \htheta \cdot \hphi ) = 0$;
finally, we take $s_a$ to be the spacelike unit-norm vector orthogonal to $u_a$, $\htheta_a$ and $\hphi_a$, which is unique up to a sign.

The tetrad $\{u_a,\ s_a,\ \htheta_a,\ \hphi_a \}$ constitutes an orthonormal basis of the (co)tangent space.
By taking tensor products of the tetrad with itself, one can then construct bases for higher-rank tensors.
For instance, the tensor
\be\label{eq:p}
P_{ab} := g_{ab} + u_a u_b
\ee
is given simply by
\be
P_{ab} = s_a s_b + \htheta_a \htheta_b + \hphi_a \hphi_b\, .
\ee
Such tensor is particularly meaningful in this context, since it acts as a projector onto the hyperplane orthogonal to $u_a$, in the following sense: for any vector $X_a$, $\vb{X}_a := P_a^{\ b} X_b$ is such that $( u \cdot \vb{X}) = 0$ and $P_a^{\ b} \vb{X}_b = \vb{X}_a$ (and similarly for higher-rank tensors).

As yet another interesting example, consider the \ae{}ther's twist tensor
\be \label{eq:twist}
\omega_{ab} := \nabla_{[a} u_{b]} + u_{[a} \mathfrak{a}_{b]}\, ,
\ee
where $\mathfrak{a}_a := u^b \nabla_b u_a $ is the \ae{}ther's acceleration and the square brackets mark antisymmetrisation.\footnote{We include the factor $1/2$, i.e.~$X_{[a} Y_{b]} : = \frac{1}{2} \left( X_a Y_b - X_b Y_a \right)$.} 
As mentioned in the introduction, \cref{s:intro}, Frobenius' theorem states that such tensor vanishes identically everywhere in the spacetime if and only if $u_a$ is orthogonal to a family of codimension-one hypersurfaces.
We will thus refer to such case, already treated extensively in \cite{bhattacharyya_causality_2016}, as the hypersurface-orthogonal case.

It is relatively simple to prove that $\omega_{ab}$, which is transverse to $u^a$ by construction, satisfies
\be
\omega_{ab} \psi^b = 0\, ,
\ee
as a consequence of \cref{eq:uZAMO} and the Killing equation; 
this entails that the twist has only one independent component, $\omega_{ab} \propto s_{[a} \htheta_{b]}$, and that
\be
\omega_{ab} \chi^b = \left(s \cdot \chi \right) \left(\htheta^c \, \omega_{cb}\, s^b \right) \htheta_a\, .
\ee

\subsection{Causality and Lorentz-violating degrees of freedom}\label{s:causality}

As a consequence of the unit-norm condition, \cref{eq:unorm}, the \ae{}ther field can never vanish and is always timelike.
It thus selects a preferred time direction in the tangent space at each spacetime point --- a violation of local Lorentz boost symmetry.\footnote{Technically, the \ae{}ther also brakes rotational symmetry in all frames except the preferred one.}
Correspondingly, the span of $\{s_a,\ \htheta_a,\ \hphi_a \}$ constitutes a preferred spatial subspace of the tangent space.

The presence of the \ae{}ther field therefore suggests a new notion of causality, not directly related to the metric and its light cones.
Following \cite{bhattacharyya_causality_2016}, we describe causality in terms of curves:
we call \emph{causal} a curve whose tangent vector $t^a$ has $(u \cdot t) \neq 0$; the curve is future-directed if $(u \cdot t) < 0$ and past-directed if $( u \cdot t) > 0$.
Note that, while timelike and null curves are always causal, spacelike curves may be causal or not.
Hence, a causal curve, has a tangent vector
\be\label{eq:tan_c}
t_a \propto u_a + c v_a\, ,
\ee
where $v_a$ is a linear combination of $s_a$, $\htheta_a$, and $\hphi_a$ and specifies the curve's spatial direction, while the factor $c$ can be interpreted as a preferred-frame speed.
The proportionality sign reflects the fact that one is free to reparametrise the curve at will.
Note that such tangent vector is timelike as long as $c<1$, but remains causal even if $c>1$ --- i.e.~if the motion is superluminal.

In Einstein--\ae{}ther theory, in the absence of matter, for example, there exist five (massless) degrees of freedom, which can be classified according to their spin: 
two tensor (spin-two) polarisations, two vector (spin-one) polarisations, and one scalar (spin zero).
Each spin representation propagates with a peculiar speed that is set by the couplings of the theory and can become superluminal \cite{JacobsonEinsteinAetherWaves2004}.

Typically, as long as the equations of motion of the theory in question are of order no higher than two, all degrees of freedom behave in a way similar to this example: 
though different (massless) degrees of freedom may propagate at different speeds, all speeds are bounded from above and do not depend directly on the spatial momentum. 
In other words, the dispersion relation of these degrees of freedom is always linear.
In such cases, one can picture the causal structure as being described by causal cones centred around $u_a$ that open up more and more the higher the propagation speed.
Such causal cones can be thought of as null cones of suitably defined effective metrics --- also known as speed-$c$ metrics.

It is important to point out, however, that once an \ae{}ther vector is introduced, nothing forbids to add higher spatial derivatives to the equations of motion. 
Such extension drastically alters the dispersion relation of the degrees of freedom propagating in the theory, rendering their propagation speed momentum dependent and possibly allowing for arbitrarily high propagation speeds.
Further support to such an extension is lent by the fact that it provides a natural way to UV complete the gravitational theory \qu{\emph{à la} Ho\v{r}ava}, i.e.~by introducing higher-order operators involving derivatives of order higher than two only in the spatial directions --- thereby improving the high-energy behaviour of propagators, and ensuring power counting renormalisability, without introducing higher derivatives in time, thus avoiding the related Ostrogradsky ghosts.\footnote{Remarkably, perturbative renormalisability has also been proven almost entirely in recent works~\cite{BarvinskyRenormalizationHorava2016,BarvinskyRenormalizationGroup2019,BarvinskyBetaFunctions2022,BellorinCancellationDivergences2022,BellorinBRSTSymmetry2022,BellorinRenormalizationNonprojectable2024}.}

These remarks motivate us to introduce a toy model for describing degrees of freedom with nonlinear dispersion relations, in the form of a massless scalar field $\phi$, with negligible backreaction on the background, and satisfying the modified Klein--Gordon equation
\be\label{eq:modifiedKG}
\Box{\phi} + \sum_{j=2}^N (-1)^j\frac{\beta_{2j}}{\Lambda^{2j-2}} \triangle^j \phi = 0\, .
\ee
Here, $\Box := g^{ab} \nabla_a \nabla_b$ is the usual d'Alembert's operator, given by the spacetime metric; 
the operator $\triangle := P^{ab} \nabla_a \nabla_b$ is instead a Laplacian, defined in terms of derivatives in the directions transverse to the \ae{}ther through the projector of \cref{eq:p}.
The $\beta_{2j}$ are couplings, while the quantity $\Lambda$ is a high energy scale; the highest power $N$ of the Laplacian controls the high-energy (Lifshitz) scaling of the equation.
We assume that all the $\beta_{2j} \geq 0$ to ensure that there are no regions in the parameter space giving rise to subluminal propagation. 
We also set $\beta_{2N}=1$, which is always true up to a rescaling of $\Lambda$; 
finally, we choose units so that the speed in the limit $\Lambda \to \infty$ is $1$, although this needs not coincide with that of light.
Therefore, the scalar field $\phi$ can be thought of as a proxy of e.g.~gravitational --- metric and/or \ae{}ther --- or matter perturbations.

We will not be interested here in the full dynamics ensuing from \cref{eq:modifiedKG}. 
Rather, we shall focus on the behaviour of its solutions in the vicinity of a (candidate) universal horizon and therefore limit our analysis to the leading order in a Wentzel--Kramers--Brillouin (WKB) expansion.
That is, we write
\be \label{eq:WKBfield}
\phi = A e^{\iu S}
\ee
and assume a slowly varying amplitude $A$ with respect to a rapidly varying phase $S$.

Plugging this Ansatz into the equation of motion, \cref{eq:modifiedKG}, we get 
\be\label{eq:DR}
g^{ab} k_a k_b +  \sum_{j=2}^N \frac{\beta_{2j}}{\Lambda^{2j-2}} \left( P^{ab} k_a k_b \right)^{j} = 0\, ,
\ee
which is a mass shell condition for $k_a := \partial_a S$.
Such condition has a twofold interpretation: 
as a geometric relation identifying codimension-one constant-phase hypersurfaces, to which $k_a$ is orthogonal;
and as a Hamilton--Jacobi equation, interpreting $k_a$ as the momentum conjugate to the position of a point particle.
Leveraging this dichotomy, we thus refer to $k_a$ as \qu{momentum}.

Employing the tetrad of \cref{s:pframe}, we may decompose 
\be\label{eq:kpf}
k_a &= \omega u_a + k_s s_a + k_\htheta \htheta_a + k_\hphi \hphi_a \, ,
\ee
where we have defined $\omega := - ( k \cdot u)$, $k_s := (k \cdot s)$, and similarly for $k_\htheta$ and $k_\hphi$.
Hence, the quantity $\omega$ can be interpreted as \qu{energy in the preferred frame} and, similarly, $\vb{k}_a:= P_a^{\ b} k_b$ as \qu{spatial momentum in the preferred frame}.
By calling 
\be
\abs{\vb{k}}:=\sqrt{P^{ab} k_a k_b} = \sqrt{k_s^2 + k_\htheta^2 + k_\hphi^2}
\ee
the norm of the spatial momentum, we realise that the WKB equation of motion, \cref{eq:DR}, is in fact a dispersion relation giving the energy as a function of the momentum
\be\label{eq:DRpf}
\omega^2 &= \abs{\vb{k}}^2  +  \sum_{j=2}^N \frac{\beta_{2j}}{\Lambda^{2j-2}} \abs{\vb{k}}^{2j} \, .
\ee

Regarding instead the WKB equation of motion, \cref{eq:DR}, as determining constant-phase contours, we realise that a curve $x^a(\lambda)$ laying on one such contour has tangent vector
\be\label{eq:xdot}
\dot{x}^a := \dv{x^a}{\lambda} \propto \left[ u^a + \frac{\omega}{ \abs{\vb{k}} } \frac{\vb{k}^a}{\abs{\vb{k}}} \right] + \alpha_i \left( e^i \right)^a\, ,
\ee
where $\left( e^i \right)^a$ are two spacelike vectors orthogonal to both $u^a$ and $\vb{k}^a$.
Such vector is defined solely by the requirement that it be orthogonal to $k_a$, hence the coefficients $\alpha_i$ are not fixed by \cref{eq:DR}.
The proportionality sign is again due to the fact that one can reparametrise the curve at will, since there is no obvious way of choosing $\lambda$.
Note that the terms within the square brackets are in the form of \cref{eq:tan_c}, provided we identify $v_a$ with $\vb{k}_a/\abs{\vb{k}}$ and $c$ with the \emph{phase velocity} 
\be\label{eq:c_phase}
c_{\rm p} \left( \abs{\vb{k}} \right) := \frac{\omega}{\abs{\vb{k}}}\, .
\ee
I.e.~constant-phase curves are causal, in the sense outlined above, and \qu{move} with the phase velocity in the direction of the momentum.
Further note that, within our assumptions, $c_{\rm p} > 1$. 

On the other hand, regarding \cref{eq:DR} as a Hamilton--Jacobi equation means that its left-hand side plays the role of a Hamiltonian. 
Of the ensuing Hamilton's equations, one determines how the momentum evolves along the particle's trajectory $X^a(\mu)$, while the other fixes the relationship between the momentum and the vector $\dot{X}^a := \dv*{X^a}{\mu}$ tangent to the trajectory.
Together, these equations give rise to a modified geodesics equation --- see e.g.~\cite{CapassoParticleKinematics2010,SuyamaNotesMatter2010} for applications in Ho\v{r}ava gravity. 
A simple computation, sketched in \cref{a:hamiltonian}, yields
\be\label{eq:Xdot}
\dot{X}_a = \omega \left[ u_a + \dv{\omega}{\abs{\vb{k}}} \frac{\vb{k}_a}{ \abs{\vb{k}} } \right] \, ,
\ee
which is again of the form of \cref{eq:tan_c} with the identification of $v_a$ with $\vb{k}_a/\abs{\vb{k}}$ and of $c$ with the \emph{group velocity}
\be
c_{\rm g} \left( \abs{\vb{k}} \right) := \dv{\omega}{\abs{\vb{k}}} \, .
\ee
That is, the trajectories that solve Hamilton's equations are causal curves that move at the group velocity. 
Note that, in our settings, $c_{\rm g} > c_{\rm p}$.

It is worth noticing that, unlike what happens in most applications of relativistic mechanics, the conjugate momentum $k_a$ is not tangent to the trajectory that solves Hamilton's equations.
In fact,
\be
\dot{X}_a &= c_{\rm g} c_{\rm p} k_a - \omega \left( c_{\rm g} c_{\rm p} - 1 \right) u_a 
\ee
has a nonvanishing component orthogonal to $k_a$:
considering the linear combination $\abs{\vb{k}} u_a + \omega v_a$, which is orthogonal to $k_a$ by construction, it is immediate to see that
\be\label{eq:ell}
 \abs{\vb{k}} \left( u_a \cdot  \dot{X} \right) + \omega \left( v \cdot  \dot{X} \right) = \omega \abs{\vb{k}} \left( c_{\rm g} c_{\rm p} - 1 \right) \neq 0\, .
\ee
This is a direct consequence of the fact that the Hamiltonian deriving from \cref{eq:DR} is a nonhomogeneous function of the momentum.
This seemingly innocuous feature is extremely consequential and leads naturally to the introduction of a notion of curvature for the momentum space (see e.g.~\cite{Girelli:2006fw,Amelino-Camelia:2011lvm,Barcaroli:2015xda,Lobo:2016xzq,Carmona:2019fwf,Relancio:2020rys} and references therein).

It is also important to point out that neither $c_{\rm p}$ nor $c_{\rm g}$ directly inform about causal relations.
Rather, by simple analytical arguments \cite{Fox_1970,BabichevKEssenceSuperluminal2008}, one can prove that no signal can propagate outside the causal cone defined by the so-called \emph{signal velocity} $c_{\rm s}$
\be
 c_{\rm s} := \lim_{\omega \to \infty} c_{\rm p} \, .
\ee
In our setting, $\omega \to \infty$ requires $\abs{\vb k} \to \infty$ and implies $c_{\rm p} \to \infty$ as well as $c_{\rm g} \to \infty$.
Hence, the limit above is infinite and the pervious statement reflects the intuition according to which nonlinear superluminal dispersion relations entail that causal cones \qu{open up} to the plane orthogonal to $u_a$. 

It is worth noting that, in this limit, the component of the momentum that is diverging more rapidly is the one along $u_a$ and therefore, in this sense, $k_a$ aligns with $u_a$ up to subleading terms.
Moreover, the vector $\dot{x}^a$ tangent to a constant-phase contour becomes
\be
\dot{x}^a \sim u^a + c_{\rm p} \frac{\vb k^a}{ \abs{\vb k}} \, ,
\ee
since the terms $\propto (e^i)_a$, which are finite, can be neglected.
This ensures that $( k \cdot \dot{x} ) = 0$, as it must by definition, and hence entails that the constant-phase contours are orthogonal to the \ae{}ther --- to leading order.
Actually, the solutions of Hamilton's equations also become orthogonal to the \ae{}ther, although one can show that for superluminal dispersions relations, possibly more general than the one we consider here, whenever $\omega$ diverges the following inequality holds strictly
\be \label{eq:cg>cp}
\lim_{\omega \to \infty} \frac{c_{\rm p}}{c_{\rm g}} < 1 \,
\ee
--- see \cite{DelPorro_2024_tesi} for a complete derivation. 
The importance of the inequality \eqref{eq:cg>cp} can be summarised as follows: although the point-particle description provided by the Hamiltonian analysis would tell us that particles propagate with the group velocity $c_{\rm g}$, the high-frequency limit \emph{must} be governed by the signal velocity. 
The relation \eqref{eq:cg>cp} shows that there cannot be any causal connection between two events connected by the trajectory \eqref{eq:Xdot}, in the limit $\omega \to \infty$, since that would lie outside the causal cone defined by $c_s$. 
This fact will be of particular relevance in \cref{s:radiative} while taking into account the radiative properties of the universal horizon.

\subsection{Conserved quantities}\label{s:cons}

The symmetries of the background give rise to two integrals of motion \cite{cropp_ray_2014,cropp_phdthesis}.
Indeed, since $\chi^a$ and $\psi^a$ are Killing and Lie-drag $u^a$, the operators $\chi^a \, \partial_a$ and $\psi^a \, \partial_a$ commute with the linear differential operator defining the modified Klein--Gordon \cref{eq:modifiedKG} and the two quantities
\be
\chi^a \, \partial_a \phi \qq{and} \psi^a \, \partial_a \phi
\ee
are therefore conserved on shell.
By examining the action of the two operators on a plane-wave solution, one realises that these quantities are nothing but the Killing energy and the component of the (Killing) angular momentum along the axis of symmetry. 

In the specific solution found in \cite{franzin_kerr_2023}, the background metric also admits a Killing tensor $K_{ab}$, which by definition satisfy the equation $\nabla_{(a}K_{bc)}=0$.
Such tensor defines an operator, whose action on the scalar field is $\nabla_a(K^{ab} \nabla_b \phi)$, that commutes with the d'Alembertian and thus generates a quantity that is conserved along solutions of the ordinary Klein--Gordon equation: the so-called Carter constant.

However, this operator does not commute with the whole modified Klein--Gordon operator (cf.~\cref{a:hamiltonian}) and it is not clear whether it could be generalised to one that would.
Indeed, there is no reason to expect that a third constant of motion will exist on a generic background different from that of \cite{franzin_kerr_2023}.
As a consequence, the orbits we shall consider in the following will not be completely integrable.

At leading WKB order, the conserved quantities can be written as
\be
-\Omega &:= \left( k \cdot \chi \right) \, \\
m &:= \left( k \cdot \psi\right) \, ,
\ee
with $\Omega$ and $m$ two real numbers.
For a thorough analysis of conserved quantities in the context of modified dispersion relations, expressed in the language of Hamiltonian mechanics, we refer the reader to \cref{a:hamiltonian}.
When expressed in terms of the tetrad of \cref{s:pframe}, the conservation equations become
\be
- \Omega &= \omega \left( u \cdot \chi \right) + k_s \left(s \cdot \chi \right) + k_\hphi \left( \hphi \cdot \chi \right) \, , \label{eq:Epf} \\
m &= k_\hphi \sqrt{\left( \psi \cdot \psi \right)} \, ; \label{eq:Lpf}
\ee
These two equations, along with the dispersion relation \cref{eq:DRpf}, constitute the system that plays the role of equations of motion.

\section{Universal horizons}\label{s:UH}

We are now in a position to characterise universal horizon in the absence of hypersurface orthogonality.
First of all, we shall lay down a set of necessary and sufficient conditions for a codimension-one hypersurface $\mathcal{S}$ to be a universal horizon.
Since the possibility of the existence of a surface satisfying such conditions is not obvious, we shall also provide a plausibility argument in this sense.
We shall then recall some details on the stealth Kerr solution of Einstein--\ae{}ther theory described in \cite{franzin_kerr_2023}, and clarify some properties of its \qu{quasi} universal horizon.
Finally, we will present a deformation of the stealth Kerr solution that does present a universal horizon.

\subsection{Quasilocal characterisation beyond hypersurface orthogonality}\label{s:characterisation}

In light of the discussion by the end of \cref{s:causality}, it should be clear that a universal horizon must be a surface to which the \ae{}ther $u_a$ is normal.
Moreover, given the symmetries, it is reasonable to assume that the universal horizon should be left invariant by the action of time translations.
This means that $\chi^a$ should be among the generators of the universal horizon.
The combination of these two requirements suggests that a universal horizon must be a hypersurface $\mathcal{S}$ such that
\be\label{eq:uchi}
\eval{\left( u \cdot \chi \right)}_\mathcal{S} = 0\, . 
\ee
(The same reasoning applies equally well to $\psi^a$, but we have assumed that $( u \cdot \psi) = 0$ everywhere.)

In the hypersurface-orthogonal case, this condition is necessary and also sufficient \cite{bhattacharyya_causality_2016}, provided it is complemented by the technical requirement that
\be\label{eq:achi}
\eval{ \left( \mathfrak{a} \cdot \chi \right)}_\mathcal{S} \neq 0\, .
\ee
In fact, \cref{eq:uchi} is a (quasi)local characterisation of the universal horizon, in the sense that away from $\mathcal{S}$ one has $( u \cdot \chi) \neq 0$ and therefore \cref{eq:uchi} can be used to determine the universal horizon's location within a given metric--\ae{}ther configuration.

If hypersurface orthogonality is not assumed, however, \cref{eq:uchi} is no longer sufficient, for a reason partially explored already in \cite{franzin_kerr_2023}:
in essence, the presence of a nonzero twist introduces a misalignment between the normal to this surface and the \ae{}ther.
To see this, consider the normal vector to the surfaces $( u \cdot \chi) = \text{const.}$
\be
n_a := \nabla_a \left( u \cdot \chi \right)\, .
\ee
In our basis, $n_a$ can be decomposed as
\be\label{eq:npf}
n_a = - \left( \mathfrak{a} \cdot \chi \right) u_a + \left( u \cdot \chi \right) \left( \mathfrak{a} \cdot s \right) s_a  +\left[ \left( u \cdot \chi \right) \left( \mathfrak{a} \cdot \htheta \right) + 2 \left( s \cdot \chi \right) \left( \htheta^b \, \omega_{bc} \, s^c \right) \right] \htheta_a\, .
\ee
Clearly, on $\mathcal{S}$ all terms proportional to $(u \cdot \chi)$ vanish, but the normal still has a component orthogonal to $u_a$ that is controlled by the only nonvanishing component of the twist.

Moreover, the magnitude of the twist determines whether $\mathcal{S}$ is a timelike, spacelike or null hypersurface, since
\be\label{eq:normn}
\eval{\left( n \cdot  n \right)}_\mathcal{S} = \eval{ - \left( \mathfrak{a} \cdot \chi \right)^2 \left[ 1 - 4 \frac{ \left( \htheta^a \, \omega_{ab} \, s^b \right)^2 }{ \left( \mathfrak{a} \cdot s \right)^2} \right]}_{\mathcal S}\, ,
\ee
where we used the relation $\left( \mathfrak{a} \cdot \chi \right) = \left( s \cdot \mathfrak{a} \right) \left( s \cdot \chi \right)$.
Hence, if the twist vanishes, $\mathcal{S}$ is surely spacelike; 
but, as the nonvanishing component of the twist grows in magnitude, $\mathcal{S}$ can become null, first, and then timelike.

On the other hand, if the twist vanishes \emph{at least locally} on $\mathcal{S}$, then this surface is truly a universal horizon.
We thus resolve to extend the quasilocal characterisation introduced in \cite{bhattacharyya_causality_2016} to the following: 
a hypersurface $\mathcal{S}$ such that
\be\label{eq:defUH}
\eval{\left( u \cdot \chi \right)}_\mathcal{S} = 0 \, , \quad
\eval{\left( \mathfrak{a} \cdot \chi \right)}_\mathcal{S} \neq 0\, , \qq{and}
\eval{ \omega_{ab} }_\mathcal{S} = 0
\ee
is a universal horizon.
In writing \cref{eq:defUH} we want to stress that the vanishing of the twist need only happen on the surface $\mathcal{S}$, while the flow of $u_a$ could be allowed to be globally not hypersurface orthogonal.

We wish to point out that while this characterisation has never been spelled out explicitly --- al least to our knowledge ---, the gist of it is known to practitioners in the field.
For instance, \cite{adam_rotating_2021} analysed numerically rotating black hole solutions in Einstein--\ae{}ther theory and used the nonvanishing of the twist to argue that those black holes do not possess universal horizons.
Still, \cref{eq:defUH} --- along with the decomposition of the normal vector \cref{eq:npf} --- is useful in that it allows to further investigate the properties of the surface $\mathcal{S}$ and their relationship with the twist tensor.
To achieve this goal, we resort to the test scalar field introduced in \cref{s:causality}, assuming the WKB approximation holds.

\subsection{Scalar field propagation}

We start by focusing on the momentum $k_a$ and compute its projection along the normal vector \cref{eq:npf}.
We obtain
\be
\left( k \cdot n \right) &= \omega \left( \mathfrak{a} \cdot \chi \right) + k_s \left( u \cdot \chi \right) \left( \mathfrak{a} \cdot s \right) \nonumber\\
&\phantom{=} + k_\htheta \left[ \left( u \cdot \chi \right) \left( \mathfrak{a} \cdot \htheta \right) + 2 \left( s \cdot \chi \right) \left( \htheta^a \, \omega_{ab}\, s^b \right) \right] \, ,
\ee
or, on $\mathcal{S}$,
\be\label{eq:kdotn}
\eval{\left( k \cdot n \right)}_\mathcal{S} = \eval{ \omega \left( \mathfrak{a} \cdot \chi \right) \left[ 1 + 2 \frac{k_\htheta}{\omega} \frac{\htheta^a\, \omega_{ab}\, s^b}{\left( \mathfrak{a} \cdot s \right)} \right] }_\mathcal{S} \, .
\ee
We note that
\be
\abs{\frac{k_\htheta}{\omega} } \leq \frac{1}{c_{\rm p}} \leq 1\, ;
\ee
moreover,
\be
\abs{2  \frac{\htheta^a\, \omega_{ab}\, s^b}{ \left( \mathfrak{a} \cdot s \right) }}_\mathcal{S} \leq 1 
\Leftrightarrow
\eval{\left( n \cdot n \right)}_\mathcal{S} \leq 0\, ;
\ee
the equalities here hold, respectively, in the far infrared ($\Lambda \to \infty$) and when $\mathcal{S}$ is null.

We thus deduce that, as long as $\mathcal{S}$ is spacelike, the sing of $\eval{( k \cdot n)}_\mathcal{S}$ is fixed by the prefactor $\omega \eval{( \mathfrak{a} \cdot \chi ) }_\mathcal{S}$, which is the same as in the hypersurface-orthogonal case.
More specifically, one can easily convince oneself that WKB modes with positive energy in the preferred frame, $\omega>0$, generically have momenta pointing \qu{inwards} on $\mathcal{S}$ --- this irrespective of the value of the twist, unless this becomes so large that $\mathcal{S}$ stops being spacelike.
Whether this fact has any implication in regards to energy fluxes is not clear.

Next, consider a curve of constant phase $x^a(\lambda)$ that crosses $\mathcal{S}$ at some point.
Using \cref{eq:xdot}, the projection of its tangent vector along the normal to $\mathcal{S}$ is
\be\label{eq:xdotn}
\eval{\left( \dot{x} \cdot n \right) }_\mathcal{S} &\propto \eval{ \left( \mathfrak{a}\cdot \chi \right) \left[ 1 + c_{\rm p} \frac{k_\htheta}{\abs{\vb{k}}} \frac{2\left(\htheta^a\, \omega_{ab} \, s^b \right)}{\left( \mathfrak{a} \cdot s \right)} + \alpha_i \left( e^i \cdot \htheta\right) \frac{2\left(\htheta^a \, \omega_{ab} \, s^b \right)}{\left( \mathfrak{a} \cdot s \right)} \right] }_\mathcal{S}\, .
\ee
As in \cref{s:causality}, this relation is not an equality because of the freedom to reparametrise the curve arbitrarily.
Still, if we choose a parametrisation so that the curve is future directed in the sense of \cref{s:causality}, then in the case that $\eval{\omega_{ab}}_\mathcal{S} = 0$ the curve is necessarily ingoing with respect to $\mathcal{S}$.
Conversely, if any amount of twist is present, no matter how small, this is no longer true: at that point on $\mathcal{S}$, there exist both ingoing and outgoing constant-phase curves.
Note that this statement does not depend on $c_{\rm p}$, as it would hold even in the hypothetical case $c_{\rm p} = 0$ --- which is actually forbidden by our dispersion relation \cref{eq:DRpf}.
This is another way of understanding why a universal horizon cannot be defined in terms of the phase velocity alone.

Finally, let us consider the trajectory $X^a(\mu)$ of a particle in the Hamiltonian picture, and again compute the projection of its tangent vector along the normal to $\mathcal{S}$:
\be\label{eq:Xdotn}
\eval{\left( \dot{X} \cdot n \right) }_\mathcal{S} = \omega \left( \mathfrak{a}\cdot \chi \right) \left[1 + c_g \frac{k_\htheta}{\abs{\vb{k}}} \frac{2\left(\htheta^a \, \omega_{ab} \, s^b \right)}{\left( \mathfrak{a} \cdot s \right)} \right]\, .
\ee
From this expression, it is clear that the future-directed causal particles that exit $\mathcal{S}$ are those whose momentum has a nonvanishing component along $\htheta$ and that move sufficiently fast, namely with 
\be\label{eq:cg_escape}
c_{\rm g} > \abs{\frac{\left( \mathfrak{a} \cdot s \right)}{2\left(\htheta_a \, \omega_{ab} \, s^b \right)}} \sqrt{1 + \left( \frac{k_s}{k_\htheta}\right)^2 + \left( \frac{k_\hphi}{k_\htheta}\right)^2 } \,.
\ee
The right-hand side attains its minimum for $k_s = k_\hphi = 0$, hence the slowest-moving particles that can escape have momentum directed only along $\htheta_a$; 
note that the speed of such particles is superluminal if $\mathcal{S}$ is spacelike.
Further note that a particle whose motion is \qu{purely $s$-wards}, in the sense that $k_\htheta = k_\hphi = 0$, can never escape.

\bigskip

The upshot of this discussion is that one can, formally, define universal horizons even if global hypersurface orthogonality cannot be guaranteed. 
Such definition is given in terms of a set of necessary conditions, and it is therefore not clear that these conditions can be met in practice.

Indeed, it is quite immediate to formulate an objection to the existence of such horizons, based on the Raychaudhuri equations.
In the following, we shall first of all elaborate on this point and show why the objection does not lead to a no-go.
Then, we shall construct a rather explicit instance of universal horizon in a non-hypersurface-orthogonal setting, based on the stealth Kerr solution of \cite{franzin_kerr_2023}.
Before moving on, however,  three comments are in order.

The first comment concerns the quasilocal characterisation of universal horizons, in the hypersurface-orthogonal case and beyond.
As mentioned, our approach is largely derived from that of \cite{bhattacharyya_causality_2016} and makes extensive use of the spacetime symmetries.
However, there exists another, relatively independent approach developed in \cite{carballo-rubio_geodesically_2022,maciel_quasilocal_2016} and based on the analysis of the expansions of two suitably defined congruences, one timelike and one spacelike:
Within this approach, one works in analogy with general relativity to define \emph{universal trapping horizons} --- identified, roughly speaking, as the locus of point at which one of the two expansions vanishes ---, hence not restricting oneself to stationary situations. 
Clearly, the two approaches should agree when their domains of applicability overlap --- and indeed they do in all known examples, as one can easily check.
It would therefore be extremely tempting to extend the second approach straightforwardly to the non-hypersurface-orthogonal case.
Our preliminary exploration suggests that some care is needed in spelling out precisely the assumptions needed and, for this reason, we resolve to defer the matter to further investigations.

A second, very brief comment concerns the stability of universal horizons in the absence of global hypersurface orthogonality.
Indeed, the requirement that the twist vanishes locally seems to suggest that the notion of universal horizon, as we have introduced it above, might be particularly fragile.
In a setting as that of Einstein--\ae{}ther theory, in which the \ae{}ther is completely dynamical, for example, one expects that generic perturbations will carry a nonzero twist.
Hence, a small \qu{\ae{}ther wave} might easily turn a spacetime exhibiting a universal horizon into one with a surface which, for instance, still satisfies \cref{eq:uchi,eq:achi} but on which the twist does not vanish.
Though this concern is real, it pertains mostly to the dynamics of the theory in question and for this reason we leave a thorough analysis of the issue for future work.
Incidentally, we point out that even in the setting of khronometric theory, in which the \ae{}ther is always hypersurface orthogonal by construction, the dynamical stability of universal horizons is far from established.

The previous paragraph brings about a third comment, concerning the nature of spacelike surfaces satisfying \cref{eq:uchi,eq:achi} but on which the twist does not vanish.
Our analysis suggests that, although they are not horizons in the causal sense, these surfaces still display interesting properties.
In particular, the fact that on one such surface the momentum can only be directed \qu{inwards} is especially intriguing.
For this reason, we deem it worth to adopt the terminology introduced in \cite{franzin_kerr_2023} and will refer to such surfaces as \emph{quasi universal horizon}.

\subsection{A plausibility argument for the existence of universal horizons}\label{s:Raychaudhuri}

As mentioned above, it is not at all clear that a surface $\mathcal{S}$ satisfying the conditions of \cref{eq:defUH} could exist.
One possible objection relies on the kinematical constraints hidden in the Raychaudhuri Raychaudhuri equations \cite{Poisson_2004}. 
Given a timelike congruence, described by $u^a$, it is always possible to decompose the tensor $B_{ab} := \nabla_a u_b$ in the hydrodynamical invariants
\be \label{eq:Bab_dec}
\nabla_a u_b = \frac{1}{3} P_{ab} \vartheta + \sigma_{ab} + \omega_{ab} - u_a \mathfrak{a}_b \,,
\ee
where $\omega_{ab}$ corresponds to the twist tensor given in \cref{eq:twist}, the shear $\sigma_{ab}$ is the symmetric-traceless part of $B_{ab}$ and $\vartheta=\nabla_a u^a$ is the expansion of the congruence (see appendix \ref{a:Raychaudhuri} for more details).

The evolution of $B_{ab}$ along $u^a$ can be rewritten in terms of $\{ \vartheta, \, \sigma_{ab},\, \omega_{ab},\, \mathfrak{a}_a \}$ and their derivatives, giving rise to kinematical constraints and to possible obstructions to the existence of a universal horizon without hypersurface orthogonality. 
An example, where this kind of mechanism is explicitly realised, is the case of a geodesic congruence ($\mathfrak{a}_a=0$). 
In this case, one can write down an evolution equation for the twist \cite{Poisson_2004}:
calling $\dot{\omega}_{ab} := u^c \nabla_c \omega_{ab}$, one has
\be \label{eq:twist_geodesic}
\eval{\dot{\omega}_{ab} }_{\mathfrak{a}=0} =  - \frac{2\vartheta}{3}\omega_{ab} - \sigma_a^{\ c} \omega_{cb} - \omega_a^{\ c} \sigma_{cb} \,,
\ee
which is a homogeneous equation in $\omega_{ab}$ without any external source. 
If $\omega_{ab}=0$ on a codimension-one hypersurface orthogonal to $u^a$, then the only possible solution admitted by \cref{eq:twist_geodesic} is $\omega_{ab}=0$ along the whole flow. 
Note that this result is independent from the geometry of the spacetime, namely it holds for any metric $g_{ab}$.

However, the same reasoning does not hold for nongeodesic congruences. 
For an accelerated \ae{}ther, the vector $\mathfrak{a}_a$ enters nontrivially the Raychaudhuri equations, acting as a source. 
The set of equations can be written down accordingly with $\mathfrak{a}_a \ne 0$ (see appendix \ref{a:Raychaudhuri}) and \cref{eq:twist_geodesic} must be modified into
\be \label{eq:twist_acc}
\dot{\omega}_{ab} = P _{[a}^{\ c}\nabla_c \mathfrak{a}_{b]} 
+ \frac{\vartheta}{3} u_{[a} \mathfrak{a}_{b]}+ u_{[a} \sigma_{b]c} \mathfrak{a}^c - u_{[a} \omega_{b]c} \mathfrak{a}^c  - \frac{2\vartheta}{3}\omega_{ab} - \sigma_a^{\ c} \omega_{cb} - \omega_a^{\ c} \sigma_{cb} \,.
\ee
Here, one can clearly see that $\mathfrak{a}_a$ acts as a source to modify the evolution of the twist so that the resulting equation is no more homogeneous, nor independent of $g_{ab}$. 
As we will show later, the kinematical obstruction given in the geodesic case is not present any more, making it possible for the twist tensor to vanish on a single hypersurface orthogonal to the \ae{}ther. 

Conceptually, the difference between the two cases lies in the dynamical requirement ($\mathfrak{a}_a=0$ for a geodesic flow): 
the kinematical relations given by the Raychaudhuri equations do not fully determine the evolution of the hydrodynamical invariants. 
That is obvious, since not all the congruences determine the same flows, and their behaviour requires some dynamical input. 
This lack of information reflects on the fact that it is not possible to fully determine the evolution of $\mathfrak{a}_a$, as explained in appendix \ref{a:Raychaudhuri}. 
Therefore, keeping $\mathfrak{a}_a$ unconstrained, in the following we will show explicitly that it is possible to build a universal horizon without hypersurface orthogonality.

\subsection{The stealth Kerr's quasi universal horizon}\label{s:stealthKerr}

In order to make the discussion self contained, we shall first recall some facts on the stealth Kerr solution of \cite{franzin_kerr_2023}.
Such solution has been constructed by focusing on a specific corner of the parameter space of Einstein--\ae{}ther theory in which three out of the four coupling constants are set to zero;\footnote{The \qu{minimal} Einstein--\ae{}ther theory corresponding to such corner is highly reminiscent of the minimal khronometric theory discussed in \cite{FranchiniRelationGeneral2021} and has received some attention lately \cite{GursesMinimalEinsteinAether2024}, since it appears to be phenomenologically indistinguishable from general relativity.} and by demanding that the metric be that of Kerr.
Consistently solving the equations of motion yields a family of solutions, only one of which is the primary focus of \cite{franzin_kerr_2023} and here.

We adopt Boyer--Lindquist coordinates $t$, $r$, $\theta$, and $\varphi$, which are adapted to the Killing symmetries in the sense that
\be
\chi^\mu = \delta^\mu_{\ t}\, , \quad \psi^\mu = \delta^\mu_{\ \varphi} \, , \qq{and} \htheta^\mu \propto \delta^\mu_{\ \theta}\, .
\ee
The solution then reads
\be \label{eq:gStealth}
\dd s^2 &= - \left( 1 - \frac{2Mr}{\Sigma} \right) \dd{t^2} - \frac{4 M a r\sin^2\theta}{\Sigma} \dd{t} \dd{\varphi} + \frac{\Sigma}{\Delta} \dd{r^2}  + \Sigma \dd{\theta^2} +\frac{A \sin^2\theta}{\Sigma} \dd{\varphi^2} \, , \\
\label{eq:uStealth}
u_\mu \dd{x^\mu} &= \mp \sqrt{\frac{ \Sigma \Delta-  \Sigma_\quh \Delta_\quh}{A}} \dd{t} - \frac{\sqrt{- \Sigma_\quh \Delta_\quh}}{\Delta} \dd{r}\, ,
\ee
where 
\be \label{eq:BL_functions}
\Sigma := r^2 + a^2 \cos^2\theta\, , \quad \Delta := r^2 + a^2 - 2Mr \, , \quad
A := (r^2 + a^2)^2 - \Delta a^2 \sin^2\theta\,  
\ee
and $M$ and $a$ (with $a^2 \leq M^2$) are respectively the mass and spin parameter.
$\Delta_\quh$ and $\Sigma_\quh$ are nothing but the functions $\Delta$ and $\Sigma$, evaluated on a surface $r = r_\quh(\theta)$ defined implicitly by the condition $\partial_r(\Sigma\Delta)=0$ (see~\cite{franzin_kerr_2023} for details), which takes the form
\be\label{eq:rquh}
\left(r_\quh - M \right) \Sigma_\quh + r_\quh \Delta_\quh = 0\, .
\ee
This is a cubic equation in $r_\quh$ and therefore it has multiple roots; however it is understood that one should always pick the largest real one.
Finally, the upper sign in the \ae{}ther solution is to be taken for $r\geq r_\quh$ while the lower sign refers to $r < r_\quh$; this choice ensures that the derivatives of the \ae{}ther are continuous at $r=r_\quh$ and that the usual boundary condition of \cref{eq:asyflat} is satisfied.

The product $(u \cdot \chi)$ coincides with the component $u_t$, but this, by construction, vanishes on $r=r_\quh$ (see \eqref{eq:uStealth}), so we can deem such surface a candidate universal horizon.
However, the solution exhibits a nonzero twist there
\be
\eval{\htheta^\mu \, \omega_{\mu \nu}\, \chi^\nu}_{r = r_\quh} &= \frac{\partial_\theta u_t}{2\sqrt{g_{\theta \theta}}} \, ,
\ee
and therefore it is certainly not a universal horizon.

Still, it might constitute a quasi universal horizon.
Since our definition is limited to spacelike surfaces, in order to ascertain this, one only needs to check whether the norm of the normal is negative.
That the answer is in the affirmative was already established in \cite{franzin_kerr_2023} by evaluating $(n \cdot n)$ numerically as a function of $\theta$, for several choices of the spin $a$.
We can now provide an analytical proof of this statement, but since most details are rather tedious and unnecessary to the present discussion we defer them to \cref{a:stealthKerr} and only report some key results.

According to our general analysis of \cref{s:characterisation}, the norm of $n_\mu$ is negative if the following inequality holds
\be
2\abs{\frac{\htheta^\mu \, \omega_{\mu \nu} \, \chi^\nu}{ \left( \mathfrak{a} \cdot \chi \right)} }_{r=r_\quh } < 1 \, 
\ee
(assuming $(\mathfrak{a} \cdot \chi)_{r_\quh} \neq 0$).
On this particular solution, we find
\be
2\abs{\frac{\htheta^\mu \, \omega_{\mu \nu} \, \chi^\nu}{ \left( \mathfrak{a} \cdot \chi \right)} }_{r=r_\quh } = \abs{\frac{\sqrt{g^{\theta \theta}} }{ \sqrt{\abs{g^{rr}}}} \frac{\partial_\theta u_t}{ \partial_r u_t}}_{r=r_\quh }\, .
\ee
After some computations, reported in \cref{a:stealthKerr}, one realises that 
\be
\eval{\frac{\partial_\theta u_t}{ \partial_r u_t} }_{r=r_\quh } = - \dv{r_\quh}{\theta}\, ,
\ee
where $\dv*{r_\quh}{\theta}$ has an explicit expression in terms of $r_\quh$.
Examining this expression closely, one can prove the inequality above and thus establish the nature of $r=r_\quh$. 

More precisely, one finds that $r=r_\quh$ is spacelike whenever $a \leq M$ and, therefore, in this regime the surface is a quasi universal horizon.
In the limit $a \to M$, instead, the surface becomes null and coincident with the (degenerate) Killing horizon.
Moreover, the \ae{}ther becomes hypersurface orthogonal and $(\mathfrak{a} \cdot \chi)_{r_\quh} \to 0$.
Hence, this limit gives rise to an \qu{extremal} universal horizon in the sense of~\cite{FranchiniBlackHole2017}.

As a by-product of this computation, one can provide an illustrative expression for the speed that a particle having \cref{eq:DR} as Hamilton--Jacobi equation must have in order to escape this quasi universal horizon.
According to \cref{eq:cg_escape}, we have
\be
c_g \geq \abs{\frac{\sqrt{-\Delta_\quh}}{\dv*{r_\quh}{\theta}} } \sqrt{1 + \left(\frac{k_s}{k_\htheta}\right)^2 + \left(\frac{k_\phi}{k_\htheta}\right)^2} \, .
\ee
Note that the derivative $\dv*{r_\quh}{\theta}$ vanishes at the poles, hence we expect that escaping from the poles should be impossible;
the derivative also vanishes on the equator, and an analogous conclusion should be drawn there.
More generally, the (classical) emissions of particles from a quasi universal horizon is likely to be highly modulated in the angle.
This has probably important implications for the computation of greybody factors.

\subsection{Adjusting quasi universal horizons into universal horizons}

The gist of the previous discussion is that the quasi universal horizon of the stealth Kerr solution is always spacelike for nonextremal Kerr. 
However, we have seen that the presence of a nonvanishing twist at such surface prevents it from being a proper universal horizon. 
Nonetheless, the very same discussion suggests a strategy for constructing an instance of a proper universal horizon, starting from a quasi universal horizon.
The key realisation is that
\be
\eval{\left( n \cdot \chi \right)}_\mathcal{S} = 0 
\ee
and therefore, whenever $\mathcal{S}$ is spacelike, one can construct out of $n_a$ a new \ae{}ther flow for which $\mathcal{S}$ is a universal horizon. 

Within an axisymmetric spacetime, one only needs to \qu{adjust} $u_a$ by giving it an additional component along $\htheta_a$:
\be \label{eq:tildeu}
u_a \to \tilde u_a = \frac{u_a + \upsilon \htheta_a}{\sqrt{1-\upsilon^2}} \,, \quad \mbox{with} \quad \eval{\upsilon}_{\mathcal{S}}=\eval{- \frac{\left( n \cdot \htheta \right) }{ \left( u \cdot n \right)}}_{\mathcal{S}} \,.
\ee
Since $\htheta_a$ lies in the orthogonal subspace of $\chi_a$, the roots of $ ( u \cdot \chi )$ coincide with those of $( \tilde{u} \cdot \chi )$, since
\be
\left( \tilde{u} \cdot \chi \right)=  \frac{\left( u \cdot \chi \right)}{\sqrt{1-\upsilon^2}} \,.
\ee
However, by construction one now has
\be
\eval{\frac{n_a}{\sqrt{1-\upsilon^2}}}_{\mathcal S}=\eval{\tilde n_a}_{\mathcal S}=\eval{ - \left( \tilde{\mathfrak  a} \cdot \chi \right) }_{\mathcal S} \tilde u_a \,,
\ee
with $\tilde{\mathfrak  a}_a = \tilde u^b \nabla_b \tilde u_a$ the acceleration of the adjusted \ae{}ther, and the corresponding twist $\tilde \omega_{ab}$ vanishes. 

Clearly, this construction is fairly generic as it is possible within any metric--\ae{ther} configuration fulfilling the symmetries \cref{eq:chiLiepsi,eq:uLie,eq:uZAMO}, provided $\abs{\upsilon} \le 1$ --- which requires $\mathcal S$ to be spacelike. 
Hence, not only are universal horizons plausible, in the sense of the discussion of \cref{s:Raychaudhuri}, they are also possible and can be constructed explicitly.

On the other hand, the construction is inherently local to a neighbourhood of $\mathcal{S}$, since it determines the function $\upsilon$ only there.
The character of the new field $\tilde u_a$ away from $\mathcal{S}$ is governed by the global shape of $\upsilon$:
if one wishes to interpret $\tilde u_a$ as an \ae{}ther over the whole spacetime, one might have to introduce further requirements on $\upsilon$.
For instance, asking that the \ae{}ther be aligned with the timelike Killing vector at infinity \cite{bhattacharyya_causality_2016} translates into the nontrivial condition
\be
\left( \tilde{u} \cdot \chi \right) \xrightarrow{\mathfrak{i}^0} -1 \iff \upsilon \xrightarrow{\mathfrak{i}^0}  0 \,.
\ee
Moreover, demanding that $\tilde u_a$ solves the equations of motion of whatever theory one is considering will generically introduce additional partial differential equations for $\upsilon$. 

A thorough analysis of the issue is beyond the scope of this article.
However, we shall complement the discussion by focusing once again on the explicit example of the Kerr solution in Einstein--\ae{}ther theory.

\subsubsection{Stealth Kerr with universal horizon}

In order to apply our discussion to the solution of \cite{franzin_kerr_2023}, we specify to the metric--\ae{}ther pair given in \cref{eq:gStealth,eq:uStealth}. 
In the Boyer-Lindquist coordinates introduced above,
\be
\hat \theta_\mu = \delta^\theta_{\ \mu} \sqrt{g_{\theta \theta}}=\delta^\theta_{\ \mu} \sqrt{\Sigma} \,,
\ee
where $\Sigma$ is given in \eqref{eq:BL_functions}, and the adjusted \ae{}ther is
\be
\tilde u_\mu = \frac{1}{\sqrt{1-\upsilon^2}} \left[ u_\mu + \upsilon \sqrt{\Sigma}  \delta_\mu^\theta \right] \, ;
\ee
in particular, we have
\be
\tilde u_r = \frac{u_r}{\sqrt{1-\upsilon^2}}  \,, \qquad \tilde u_\theta= \frac{\upsilon \sqrt{\Sigma}}{\sqrt{1-\upsilon^2}} \,.
\ee
As argued, the surface $r=r_\quh(\theta)$ will then be a proper universal horizon for $\tilde{u}_\mu$.

An obvious question at this point is what condition one should impose on $\upsilon$ in order for $\tilde u_a$ to still satisfy the vacuum field equations of Einstein--\ae{}ther theory --- or, rather, of the specific corner of the coupling space for which the Kerr metric is a solution.
Following \cite{franzin_kerr_2023}, the only relevant equation is
\be \label{eq:EOM_new_U}
0 &=\partial_r \left( \Delta \tilde u_r \right) + \frac{1}{\sin \theta} \partial_\theta \left( \tilde u_\theta \sin \theta \right) \nonumber \\
&= \partial_r \left(  \frac{\Delta u_r}{\sqrt{1-\upsilon^2}}  \right) + \frac{1}{\sin \theta} \partial_\theta \left( \frac{\upsilon \sqrt{\Sigma} \sin \theta}{\sqrt{1-\upsilon^2}}  \right) \,.
\ee
Using the explicit form of $u_r$ from \cref{eq:uStealth}, one can convert \cref{eq:EOM_new_U} into a first order, nonlinear, partial differential equation for $\upsilon$. 
Provided $\abs{ \upsilon } < 1$, after some simple algebra we get
\be \label{eq:PDE_upsilon}
-\Theta(\theta) \upsilon \, \partial_r \upsilon + \sqrt{\Sigma}\,  \partial_\theta \upsilon + f( r,\theta)(1-\upsilon^2) \upsilon=0 \,,
\ee
where we have defined\footnote{Note that, with respect to the convention used in \cite{franzin_kerr_2023}, here we have included the factor $M^2$ into the definition of $\Theta$.}
\be
\Theta(\theta) = \sqrt{- \Sigma_\quh \Delta_\quh} 
\qq{and} 
f(r, \theta)= \frac{1}{\sin \theta} \partial_\theta \left(\sqrt{\Sigma} \sin \theta \right) \,.
\ee
The condition that $\tilde u_a$ aligns with the normal to the surface $(\tilde{u} \cdot \chi ) = 0$ represents a boundary condition for \cref{eq:PDE_upsilon}, which becomes now a Cauchy problem that can be parametrised through the radial coordinate. 

The existence of the solution to such a problem can be studied through standard techniques \cite{perko1991differential}. 
Unfortunately, given the highly nonlinear structure of the problem, one cannot be certain of the global existence of a solution, as general theorems only ensure local existence. 

In light of this fact, one may focus on a neighbourhood of $r=r_\quh (\theta)$ by assuming an ansatz in the form
\be
\upsilon (r, \theta)= \sum_{i=0}^{\infty} c_i(\theta) \, \left[ r-r_\quh(\theta) \right]^i
\qq{with} 
c_0(\theta)=\upsilon \left( r_\quh(\theta), \theta \right) = \eval{-\frac{n_\theta g^{\theta \theta}}{u_\mu n^\mu}}_\quh \,.
\ee
One can then solve iteratively the equation for the coefficients $c_i(\theta)$, order by order. 
For this first coefficient, the equation becomes
\be
\left[ -\Theta \, c_0 + \sqrt{\Sigma_\quh} \, r'_\quh \right] c_1 = - \sqrt{\Sigma_\quh} \, c'_0 - c_0 (1-c_0^2) f_\quh
\ee
where we have omitted the $\theta$-dependence of the functions and we have indicated with a prime the derivative with respect to $\theta$. 
The existence of a solution for $c_1(\theta)$ is guaranteed provided
\be
\left[ -\Theta \, c_0 + \sqrt{\Sigma_\quh} \, r'_\quh \right] \ne 0 \,,
\ee
which corresponds exactly to the so-called \textit{compatibility condition} that ensures the local existence of the solution of our Cauchy problem \cite{perko1991differential}. 

So, the stealth Kerr solution given in \cite{franzin_kerr_2023} can be deformed, at least locally, to a solution admitting a rotating universal horizon. 
However, not much can be argued about the behaviour of the \ae{}ther far away from the horizon:
whether a vacuum, Kerr-like, solution can exist for all values of $r$, and what its asymptotic behaviour might be, remains a matter for future investigations.

\section{Radiative properties}\label{s:radiative}

In the previous section we argued why universal horizons, as we have defined them in \cref{eq:defUH}, represent causal boundaries:
in essence, we showed that causal modes can only cross such horizons inwards, since escaping them would require travelling \qu{backwards in time}.
In our computations, we have often evaluated expressions directly on the universal horizon (e.g.~in \cref{eq:kdotn,eq:xdotn,eq:Xdotn}) and, in doing so, we have implicitly assumed that those expressions were \qu{regular} on $\mathcal{S}$.
More specifically, we have assumed that the curves describing the propagation of WKB modes were of class $\mathcal{C}^1$ --- which is what one would expect for trajectories in classical physics.

In this section, we shall instead be interested in solutions to the WKB equations of motion that are \qu{singular}, namely for which the momentum is nonanalytic or discontinuous at the universal horizon.
Clearly, these solutions do not represent classical propagation through the universal horizon.
Rather, they inform on the semiclassical behaviour of these black holes.
Indeed, in the spherically symmetric (and hence hypersurface-orthogonal) case these solutions have been shown to be associated with the particle pairs associated to Hawking radiation~\cite{del_porro_hawking_2023}. We shall show in this section that, even in an axisymmetric case, the conservations of the Killing energy \eqref{eq:Epf} and the angular momentum \eqref{eq:Lpf} constraint the system enough to generate a similar phenomenon. 
Let us stress that hereinafter we will focus solely on the case in which $\mathcal{S}$ satisfies \cref{eq:defUH}, namely defining a universal horizon.

\subsection{Singular trajectories}
\label{s:singtraj}

Let us consider once again the scalar field $\phi$ satisfying \cref{eq:modifiedKG}. 
The system of \cref{eq:DRpf,eq:Epf,eq:Lpf} enforces the equation of motion within the WKB approximation --- namely, the dispersion relation --- and shows the separability of the field equation \eqref{eq:modifiedKG} from the angular variable $\varphi$ and the Killing time $t$. 
The set of its solution is therefore spanned by functions $\{ \phi_{\Omega,m} \}$ that are eigenfunctions of $i \chi^a \, \partial_a$ and $i \varphi^a \, \partial_a)$, respectively with eigenvalues $\Omega$ and $m$. 
In the WKB Ansatz, we shall write
\be \label{eq:mode_omega_m}
\phi_{\Omega,m}= \phi_0 \exp \left[ - i \Omega t + i \int (k_s s_a {\rm d }x^a + k_\theta \hat \theta_a {\rm d }x^a) - i m \varphi  \right] \,.
\ee

As we explained in \cref{s:causality}, within the WKB regime $\phi_{\Omega,m}$ admits an interpretation in terms of a particle with preferred energy $\omega(\vb{k})$ and preferred momentum $\vb k_a$, propagating over the background geometry at its group velocity.
However, as mentioned, causal relations in the near-horizon limit are better captured by studying constant-phase contours, characterised by the phase velocity $c_{\rm p}$.
This is because the curves determined by $c_{\rm p}$, in the limit of divergent momentum, describe the edges of the causal cone, which is exactly what we expect an outgoing ray to follow in the attempt to climb the gravitational potential close to the horizon. 
Mathematically speaking, the analyticity of $\phi_{\Omega,m}$ implies that, in the infinite-frequency limit $\omega \to \infty$, no propagation can happen outside that causal cone \cite{Fox_1970}. 

Keeping this in mind, we define (cf.~\cref{s:causality})
\be
v^a := \frac{\vb k^a}{| \vb k |} \,.
\ee
Note that $( u \cdot v) = 0$ by definition. 
The constant phase contours of \eqref{eq:mode_omega_m} are locally given by
\be
\dd \bar u := (c_{\rm p} u_a + v_a) \dd x^a = 0 \,.
\ee
Now, let us assume to have found a solution of the system of equations \cref{eq:DRpf,eq:Epf,eq:Lpf} in terms of a curve $x^a(\lambda)$, parametrised by $\lambda \in \mathbb{R}$, along which the phase $S$ of \eqref{eq:mode_omega_m} remains constant. 
In our orthonormal basis $\{ u^a,\, s^a,\, \hat \theta^a,\, \hat \varphi^a \}$ this implies:
\be \label{eq:constant_action}
\eval{S}_{x^a(\lambda)}= \int \rmd \lambda \left[  \omega (u \cdot \dot x) + k_s (s \cdot \dot x) +  k_{\hat \theta} (\hat \theta \cdot \dot x) + k_\varphi (\hat \varphi \cdot \dot x) \right] = {\rm const. } 
\ee
Let us now introduce a set of coordinates along each of the orthonormal direction of the basis through 
\be 
\rmd \tau(\lambda)= (u \cdot \dot x) \rmd \lambda \,, \quad \rmd \rho(\lambda)= (s \cdot \dot x) \rmd \lambda \,, \quad \rmd \theta(\lambda)= (\hat \theta \cdot \dot x) \rmd \lambda \,, \quad \rmd \varphi(\lambda)= (\hat \varphi \cdot \dot x) \rmd \lambda \,,
\ee
such that the constancy of $S$ can be locally rewritten as
\be \label{eq:trajectory_sing}
  c_{\rm p} \dot \tau + v_s \dot \rho +  v_{\hat \theta} \dot \theta + v_\varphi \dot \varphi  = \left(  c_{\rm p} u_a +  v_a  \right) \dot x^a = \dot{\bar u}=0 \,.
\ee
where the dotted quantities are differentiated with respect to the parameter $\lambda$. 
Let us now perform a change of basis, in order to take into account the variation of the Killing time coordinate along the trajectory. 
First of all, we define 
\be
\xi_a :=\frac{1}{|\chi|^2-(\chi \cdot \hphi)^2} \left[ \chi_a - (\chi \cdot \hphi ) \hphi_a \right]\,,
\ee
such that $(\xi \cdot \hphi)=0$ and $(\xi \cdot \chi)=1$. 
In a system of coordinates where $\chi^a \partial_a = \partial_t$, the one-form $\xi_a \dd x^a$ captures the direction in the cotangent space along the Killing time. 
To show that, let us consider the following function:
\be \label{eq:KT}
f_\Omega= \exp \left[ - i \Omega \int \xi_a \rmd x^a  \right] \,.
\ee
We can show that $f_\Omega$ represents a WKB eigenfunction of $\chi^a  \partial_a$ with eigenvalue $(-i \Omega)$ when $(u \cdot \chi) \to 0$:
\be
\chi^a \partial_a f_\Omega= -i \Omega (\chi \cdot \xi) f_\Omega=-i \Omega f_\Omega \,.
\ee
So, since the mode $\phi_{\Omega,m}$ satisfies the same eigenvalue equation, that must contain a factor $f_\Omega$. 
Namely, one can rewrite \cref{eq:mode_omega_m} replacing the Killing time $t$ with the an integral over $\xi_\mu \dd x^\mu \equiv \dd t$. 
Therefore, we can define
\be
\dot t := (\xi \cdot \dot x)=\frac{ (\chi \cdot u)\dot \tau+(\chi \cdot s)\dot \rho}{|\chi|^2-(\chi \cdot \hphi)^2} \,.
\ee
Equivalently one can compute the variation of $t$ with respect to $\tau$:
\be
\frac{\rmd t}{\rmd \tau}=\frac{1}{|\chi|^2-(\chi \cdot \hphi)^2} \left[- (\chi \cdot u)+(\chi \cdot s)\frac{ \rmd \rho}{\rmd \tau} \right]\, \quad \mbox{with} \quad \frac{ \rmd \rho}{\rmd \tau}=- c_{\rm p} v_s=- \frac{\omega k_s}{|\vb{k}|^2} \,.
\ee
At the leading order near $\mathcal{S}$ one can the expression above becomes
\be \label{eq:dtdtau}
\frac{\rmd t}{\rmd \tau}=-\frac{1}{(\chi \cdot s)} \frac{\omega k_s}{|\vb{k}|^2} \,.
\ee
The right hand side of the \cref{eq:dtdtau} can be further simplified through the relation \eqref{eq:Epf}. 

However, the final result depends strongly on the behaviour of $k_s$ and $k_\htheta$. 
In particular, both of them enter the expression \eqref{eq:dtdtau} and both of them are, in principle, allowed to diverge near $\mathcal S$. 
It is then quite intuitive to understand that the behaviour of the trajectory at the horizon is governed by the ratio
\be
\zeta = \frac{k_{s}}{k_{\hat \theta}} \biggr|_{\mathcal{S}} \,, \qquad \zeta \in \mathbb{R} \cup \{\pm \infty\}\,,
\ee
which parametrises the direction in the $\{ s, \, \hat \theta\}$ plane with which the surface $\mathcal{S}$ is approached by the ray. 
In principle, $\zeta$ can take values on the whole real line up to $\pm \infty$, depending on the behaviour of the spatial components of the momentum. 
Let us remind that we are assuming $|\vb{k}| \to \infty$ near $\mathcal S$, therefore at least one between $\{ k_s, \, k_\htheta \}$ has to blow up. 
In particular, a closer look to \cref{eq:DRpf,eq:Epf,eq:Lpf} tells us that $\zeta \ne 0$ implies that $k_s$ diverges and $\zeta \ne \pm \infty$ implies that so does $k_\htheta$. 
The limiting cases $\zeta=0$ and $\zeta = \pm \infty$ correspond to $k_\htheta$ or $k_s$ being the dominant component of $|\vb{k}|$, respectively.\footnote{Interestingly, one can easily see that the conservation of $\Omega$ constraints the case $\zeta=0$ to correspond to a finite $k_s$}

\subsubsection{Singular trajectories: \texorpdfstring{$\zeta \ne 0$}{ζ ≠ 0} }\label{s:kappa_peeling}

If $\zeta \ne 0$ (namely $|k_s| \to \infty$), the right hand side of \eqref{eq:dtdtau} can be strongly simplified through \eqref{eq:Epf}, at the leading order near $\mathcal{S}$, to
\be \label{eq:zeta_peeling}
\frac{\rmd t}{\rmd \tau}=  \frac{k_s^2}{|\vb k|^2} \frac{1}{(u \cdot \chi)}  =  \frac{\zeta^2}{1+\zeta^2} \frac{1}{(u \cdot \chi)} \,.
\ee
From \cref{eq:zeta_peeling} it is immediately clear why the case $\zeta=0$ represent a special case, since the leading order vanishes and the expansion has to be performed at the next-to-leading order. 
Remarkably, the expression \eqref{eq:zeta_peeling} shows a simple pole on the trajectory at $(u \cdot \chi)=  0$.

Moreover, without any loss of generality, we can choose a parametrisation of $x^a(\lambda)$ in such a way that $(u \cdot \dot x)=-1$ and $\lambda= -\tau$, thus picking up the \ae{}ther time as a well-defined parameter along the curve. 
Since the integral lines of $u^a$ intersects $\mathcal S$ at a finite value of $\tau$, say $\tau= \bar \tau$, we have:
\be \label{eq:uchi_expansion}
(u \cdot \chi)(\tau) = (\mathfrak a \cdot \chi)|_{\bar \tau}(\tau - \Bar{\tau}) + \mathcal{O}((\tau-\Bar{\tau})^2) \,.
\ee
If we plug \cref{eq:uchi_expansion} into \eqref{eq:zeta_peeling}, working only at the leading order in $(\tau-\bar \tau)$, we have
\be \label{eq:tdot_final}
\frac{\rmd t}{\rmd \tau} = \left[   \frac{\zeta^2}{1+\zeta^2}  \frac{1}{  (\mathfrak{a} \cdot \chi)}  \right]_{\bar \tau} \frac{1}{\tau - \bar \tau} \implies t = \left[  \frac{\zeta^2}{1+\zeta^2}  \frac{1}{  (\mathfrak{a} \cdot \chi)}  \right]_{\bar \tau} \log(\tau - \bar \tau) \,,
\ee
Interestingly, in approaching $\mathcal{S}$, the singular trajectories show a logarithmic behaviour along the Killing direction, as it happens in the spherically symmetric scenario \cite{del_porro_hawking_2023}. 
The prefactor
\be \label{eq:kappa_peeling}
\kappa = \eval{\frac{1+\zeta^2}{ \zeta^2 } (\mathfrak{a} \cdot \chi) }_{\mathcal{S}} \,
\ee
plays the role of the \qu{peeling surface gravity} of the universal horizon and depends on the particular trajectory that we are describing: the value of $\zeta$ exactly parametrises the direction along which the ray \qu{peels off} from $\mathcal{S}$. 
The difference with the spherically symmetric case can be interpreted as an inheritance of the underdetermination of the system \cref{eq:DRpf,eq:Epf,eq:Lpf} and the lack of a further relation linking $k_s$ and $k_{\hat \theta}$. 
We shall comment about this feature later on in this article.

As a final remark for this section, let us comment on the expression \eqref{eq:kappa_peeling}. 
First of all, we point out that $\kappa$ is proportional to $(\mathfrak{a} \cdot \chi)$, as in the hypersurface-orthogonal case \cite{cropp_surface_2013}, through a $\zeta$-dependent factor. 
The spherically symmetric limit is recovered with $\zeta^2 \to \infty$ matching with previous results \cite{del_porro_gravitational_2022,del_porro_hawking_2023}, as expected. 

Note that, in general, $(\mathfrak{a} \cdot \chi)$ (and so $\kappa$) will not be constant on $\mathcal{S}$ along the $\hat \theta$-direction.  
However, if the only nonvanishing component of the twist $\omega_{ab}$ has a vanishing normal derivative, namely
\be \label{eq:constant_kappa}
  u^a \partial_a (\hat \theta^b \omega_{bc} \chi^c)|_{\mathcal{S}}=0 \iff \hat \theta^a \partial_a (\mathfrak{a} \cdot \chi)|_{\mathcal{S}}=0 \,,
\ee
then $(\mathfrak{a} \cdot \chi)$ is constant (but still $\zeta$-dependent) on $\mathcal{S}$. 
For a complete derivation of \cref{eq:constant_kappa}, we refer to appendix \ref{a:Raychaudhuri}. 
We shall comment on the implication of this relation later on on this article

\subsubsection{Singular trajectories: \texorpdfstring{$\zeta = 0$}{ζ = 0}}\label{s:zetazero}

A rather special analysis has to be reserved to the case in which $\zeta=0$, that is when $k_s$ remains finite and only $k_\htheta$ diverges. 
In this case the expansion \eqref{eq:zeta_peeling} must be computed at the next order:
\be 
\frac{\rmd t}{\rmd \tau} = \frac{ \Xi k_s^2 }{(u \cdot \chi) k_{\hat \theta}^2} \,, \qquad \Xi=\frac{\Omega+ k_s (s \cdot \chi) + k_\varphi (\hat \varphi \cdot \chi)}{ (s \cdot \chi) k_s} \,.
\ee
We can evaluate the divergence of $\dv*{t}{\tau}$ in this case just by solving the system \cref{eq:DRpf,eq:Epf,eq:Lpf} for a finite $k_s$ and a divergent $k_{\hat \theta}$ which, at the leading order gives
\be \label{eq:tdot_integrable}
k^2_{\hat \theta} \propto \frac{1}{(u \cdot \chi)^{\frac{2}{N}}} \implies \frac{\rmd t}{\rmd \tau} = \frac{ \tilde \Xi}{(u \cdot \chi)^{1-\frac{2}{N}}}\,
\ee
where we collected all the finite terms in $\tilde \Xi$. 
Remarkably, $\dot t$ does not display a simple pole any more;
rather, it diverges as a power law with an exponent which, by definition, respects the following bound
\be 
0 \le 1-\frac{2}{N} < 1\,.
\ee
This implies that $\dv*{t}{\tau}$ still diverges but its expression \eqref{eq:tdot_integrable} represents an integrable singularity at $\bar \tau$ and $t$ approaches a constant value on $\mathcal S$:
\be 
t(\tau) =t(\bar \tau)+ \frac{N}{2} \tilde \Xi  (\mathfrak{a} \cdot \chi)^{\frac{2}{N}-1} (\tau - \bar \tau)^{\frac{2}{N}} \,.
\ee

Therefore, for $\zeta=0$, the singular trajectories approach $(u \cdot \chi)=0$ without any logarithmic divergence.\footnote{Actually, the trajectory appears to be $\mathcal C^0$. Still the attribute \qu{singular} will be maintained, due to the nondifferentiability of the curve.}
This has a crucial consequence on the validity of the WKB approximation for the modes with $\zeta=0$. 
We shall address this problem in the next section.

\subsection{Adiabaticity of the singular modes} \label{s:adiabaticity}

The outcome of the previous sections is clear: 
two different types of singular trajectories arise from \cref{eq:DRpf,eq:Epf,eq:Lpf}, depending on the values of $\zeta$. 
Near $\mathcal{S}$, the behaviour of the rays can be summarised by the following formula:
\be \label{eq:ubar_singular}
 \rmd \bar u=\rmd t - \mathcal{A} \frac{ \rmd \tau}{ (\tau-\bar \tau)^\alpha} =0 \,,
\ee
where, we remind the reader, $\bar u$ is the effective outgoing coordinate followed by a particle travelling on the constant phase contours of \eqref{eq:mode_omega_m}.
In the case $\zeta \ne 0$ we identify
\be 
 \alpha=1 \,, \qquad \mathcal{A}= \frac{1}{\kappa} \,,
\ee
while for $\zeta=0$
\be 
 \alpha=1-\frac{2}{N} \,, \qquad \mathcal{A}= \tilde \Xi  (\mathfrak{a} \cdot \chi)^{\frac{2}{N}-1} \,.
\ee

Before moving on to the implication that \cref{eq:ubar_singular} has on the radiative properties of $\mathcal S$, let us make an observation. 
Our trajectories have been derived determining the constant phase contours of \cref{eq:mode_omega_m}, which is intended to be a solution the system \cref{eq:DRpf,eq:Epf,eq:Lpf} within the WKB approximation. 
However, if this approximation fails, the field equation \eqref{eq:modifiedKG} cannot be mapped into the dispersion relation \eqref{eq:DRpf} and we cannot trust \cref{eq:ubar_singular} as the equation for the constant-phase contours of $\phi_{\Omega,m}$. 
In particular, the WKB field has to respect the \emph{adiabaticity condition}:
\be \label{eq:adiabaticity_condition}
 \left| \frac{\dot k_a}{k^2_a} \right| \ll 1 \,.
\ee
In other words, as long as we can neglect the variation of $k_a$ with respect to $k_a$ itself, \cref{eq:modifiedKG} coincides with \cref{eq:DRpf}.

The condition \eqref{eq:adiabaticity_condition} involves a time derivative of the component of $k_a$ in the numerator and it is therefore an observer-dependent statement. 
Within a general-relativistic treatment, this condition is usually evaluated referring to the proper time of a freely infalling observer \cite{Dey_2019_quantum_atmosphere}. 
Such observer crosses the Killing horizon smoothly and therefore experiences the exponential peel-off that characterises the outgoing rays. 
However, the prescription of free fall is not, strictly speaking, necessary in order to draw such conclusion:
any ingoing observer who crosses $\mathcal S$ smoothly will detect the adiabaticity condition for the outgoing rays to be exactly fulfilled. 

In a theory with a preferred frame, where no universal notion of free fall is prescribed\footnote{Here \qu{universal} means \qu{independent from the way the observer couples to the \ae{}ther}.}, the most natural choice is to take an observer coherent with the \ae{}ther, namely the one for which $u^a$ represents the four-velocity. 
For this trajectory, naturally parametrised by $\tau$, we have, in the basis $\{u^a,\, s^a ,\, \hat \theta^a ,\, \hat \varphi^a \}$
\be \label{eq:infalling_observer}
s_a \rmd x^a=\hat \theta_a \rmd x^a= \hat \varphi_a \rmd x^a=0 \,, \qquad u_a \rmd x^a =\rmd \tau \,.
\ee

Therefore we can compute the separation between two close-by singular trajectories \eqref{eq:ubar_singular}, as measured by someone following the world line \eqref{eq:infalling_observer}:
\be 
\frac{ \rmd \bar u}{\rmd \tau}= -  \frac{\mathcal{A}}{ (\tau-\bar \tau)^\alpha} + \cdots \,.
\ee
Equation \eqref{eq:adiabaticity_condition} has then to be understood as
\be \label{eq:adiabatic_time}
 \frac{\rmd k_a}{\rmd \bar u} \frac{1}{k^2_a} = \frac{\rmd \tau}{\rmd \bar u} \frac{\rmd k_a}{\rmd  \tau} \frac{1}{k^2_a}  \ll 1 \,,
\ee
which tells us how $k_a$ varies with respect to $k_a^2$, as detected by the ingoing observer. 
In the following, we shall check if this condition is fulfilled. 

Let us begin with $\zeta \ne 0$. In this case $\alpha=1$ and both $k_s$ and $k_\htheta$ blow up as
\be 
 \zeta k_\htheta=k_s \propto (u \cdot \chi)^{- \frac{1}{N-1}} \propto (\tau-\bar \tau)^{- \frac{1}{N-1}}\,.
\ee
For these rays, $k_\theta$, if it diverges, exhibits the same behaviour as $k_s$, as we pointed out in the previous section. 
We have therefore:
\be 
 \frac{\rmd \tau}{\rmd \bar u} \frac{\rmd k_s}{\rmd  \tau} \frac{1}{k^2_s} \propto (\tau-\bar \tau)^{1-\frac{N-2}{N-1}}= (\tau-\bar \tau)^{\frac{1}{N-1}}\xrightarrow{\tau \to \bar \tau} 0 \,.
\ee
This tells us that the adiabatic condition, due to the logarithmic behaviour of the trajectory, is fulfilled \emph{exactly}, similarly to what happens in the relativistic case for spherical symmetry \cite{del_porro_hawking_2023}. 
One can check explicitly that the same is true for $\omega$ and $k_\varphi$. 
As an important consequence, the WKB solution \eqref{eq:mode_omega_m} becomes an exact solution of \cref{eq:modifiedKG} in the near-$\mathcal S$ limit. 

However, if we analyse the case $\zeta=0$ we have $\alpha=(N-2)/N$ and $k_{\hat \theta}$ diverging as
\be 
 k_{\hat \theta} \propto (u \cdot \chi)^{- \frac{1}{N}} \propto (\tau-\bar \tau)^{- \frac{1}{N}}\,.
\ee
Plugging this behaviour into \cref{eq:adiabatic_time} we get
\be 
 \frac{\rmd \tau}{\rmd \bar u} \frac{\rmd k_{\hat \theta}}{\rmd  \tau} \frac{1}{k^2_{\hat \theta}} \propto (\tau-\bar \tau)^{1-\frac{2}{N}+\frac{1}{N}-1}=(\tau-\bar \tau)^{-\frac{1}{N}}  \,,
\ee
which evidently maximally violates the condition given in \cref{eq:adiabaticity_condition} when approaching $\mathcal S$. 

Therefore, the rays associated to the case $\zeta=0$ cannot be studied within the WKB approximation near $(u \cdot \chi)=0$, as they lose their point-particle interpretation. 
More importantly, if the field $\phi$ does not enjoy an adiabatic evolution, it is impossible to treat it as an asymptotic state and to associate a vacuum state to it.  
Therefore, rays associated to $\zeta=0$ will not represent a good tool to probe the radiative properties of $\mathcal{S}$.

\subsection{Fixing \texorpdfstring{$\zeta$}{ζ} with a kinematical argument} \label{s:fixing_zeta}

In this section, we provide an argument, based on causality, to fix the values of $\zeta$ for the irregular trajectories. 
In section \ref{s:characterisation} we have characterised the universal horizon as the surface which all the regular trajectories must cross going \qu{inwards}.
Here, \qu{inwards} means that the ray's tangent vector aligns positively with the normal $n$: $(\dot x \cdot n) >0$. 
The corresponding definition of an outgoing trajectory is then, near $\mathcal S$,
\be \label{eq:outgoing_def}
0> \dot x^a \partial_a (u \cdot \chi) = \dot{x} \cdot n \,.
\ee
In particular, we can consider one of our singular trajectories $\dot x^a(\tau)$ --- where the normalisation has again been chosen so that $(u \cdot \dot{x}) = -1$ as in \cref{s:zetazero} --- and contract it with the vector $n_a$ given in the vicinity of the universal horizon:
\be
\left( n \cdot \dot x \right) = (\mathfrak{a} \cdot \chi) + \omega \frac{\left[ k_s (\mathfrak{a} \cdot s)+k_\htheta (\mathfrak{a} \cdot \htheta) \right]  (u \cdot \chi) + 2 k_\htheta (s \cdot \chi) (\htheta^a \omega_{ab} s^b )}{|\vb k|^2} \,.
\ee
As already shown, for a regular trajectory the condition above implies $(n \cdot \dot x)=(\mathfrak{a} \cdot \chi)>0$ on $\mathcal S$, and no outgoing trajectory is allowed. 
However, for a singular trajectory, this condition can be rewritten in terms of $\zeta$, near the universal horizon, using the conservation of the Killing energy \eqref{eq:Epf}:
\be
\left( n \cdot \dot x\right) =\frac{(s \cdot \chi)}{1 + \zeta^2} \left[ (\mathfrak{a} \cdot s) - \zeta \left( (\mathfrak{a} \cdot \htheta) + 2 \frac{(\htheta^a \omega_{ab} \chi^b )}{(u \cdot \chi)} \right)\right]  \,.
\ee

Since for a true universal horizon $n_a = - (\mathfrak{a} \cdot \chi) u_a$, a trajectory satisfying the strict inequality $(n \cdot \dot x)_{\mathcal S} < 0 $ would be past directed.
We can then conclude that any singular trajectory which is not ingoing will have to satisfy
\be \label{eq:zeta_bound}
\lim_{\tau \to \bar \tau}(n \cdot \dot x)=0 \implies \zeta=\zeta_1= \eval{\frac{(\mathfrak{a} \cdot s)}{ (\mathfrak{a} \cdot \htheta) + 2 \frac{(\htheta^a \omega_{ab} \chi^b )}{(u \cdot \chi)} } }_{\mathcal S} \quad \mbox{or} \quad \zeta=\zeta_2= \pm \infty \,.
\ee
Therefore, in order to respect the causal structure of the spacetime, the ray can either travel along some fixed-$\htheta$ hyperplane $(\zeta= \pm \infty)$ or approach the universal horizon with a finite $\zeta_1$, which depends on the acceleration $\mathfrak{a}_a$. 

Let us now analyse the various terms that appear in $\zeta_1$ of \cref{eq:zeta_bound}. 
First of all, the ratio $(\htheta^a \omega_{ab} \chi^b )/(u \cdot \chi)$, which in principle represents an indeterminate form $[0]/[0]$ at the universal horizon, is actually finite and can be rewritten as (see \cref{eq:omega_dot_UH} in appendix \ref{a:Raychaudhuri})
\be
\lim_{\tau \to \bar \tau} \frac{(\htheta^a \omega_{ab} \chi^b )}{(u \cdot \chi)}= \eval{\htheta^a \partial_a \log (\mathfrak{a} \cdot \chi)}_{\mathcal S} \,.
\ee
Secondly, we stress that \cref{eq:defUH} implies $(\mathfrak{a} \cdot s) \ne 0$, from which we can see that $\zeta_1$ cannot vanish.
Hence, causality provides a second reason --- in addition to the violation of the WKB approximation --- to exclude the peculiar case $\zeta = 0$, associated to outgoing trajectories moving only along $\htheta^a$.

Note also that, by considering the limit of a spherically symmetric universal horizon, the known result is recovered: indeed in that case we have $(\mathfrak a \cdot \htheta) = (\htheta^a \omega_{ab} \chi^b ) = 0$ which straightforwardly implies $\zeta_1=\zeta_2 = \pm \infty$. 

Before closing this section, let us add an interesting observation. 
One can also rewrite $\zeta_1$ as
\be 
\zeta_1= \eval{\frac{(\mathfrak{a} \cdot s)}{ (\mathfrak{a} \cdot \htheta) + 2 \frac{(\htheta^a \omega_{ab} \chi^b )}{(u \cdot \chi)} } }_{\mathcal S}=\eval{\frac{n \cdot s}{n \cdot \htheta}}_{\mathcal S}  \,.
\ee
The advantage of this rewriting can be seen as follows: 
let us consider a vector $V^a$ orthogonal to $u^a$, $\hphi^a$ and to the spatial parts of $n^a$
\be 
\left( V \cdot u \right) = V^a P_{ab} n^b= \left( V \cdot \hphi \right) = 0 \,,
\ee
then the imposition $\zeta=\zeta_1$ is equivalent to saying that the component $(k \cdot V) = k_V$ remains finite while approaching the universal horizon. 
This fact, despite being a simple rewriting of the condition for the trajectory to be outgoing, seems to have a rather special geometrical meaning. 
In fact, we observe that $V^a$ is quite a special vector, since near $\mathcal S$ it Lie-drags the \ae{}ther field so that
\be \label{eq:V_Lie_dragging}
\mathcal L_V u_a=0+ \mathcal O (u \cdot \chi) \,.
\ee
The relation above is a direct consequence of the vanishing of the twist on $\mathcal S$. 
In fact, any one-form which is Lie-dragged by three different, linearly independent vectors is hypersurface orthogonal, and \cref{eq:V_Lie_dragging} highlights exactly the approximate hypersurface orthogonality of our \ae{}ther.

\subsection{Radiation from rotating UHs: a tunneling computation}

The arguments provided in sections \ref{s:fixing_zeta} and \ref{s:adiabaticity} point in the same direction: the irregular trajectories are characterised by a logarithmic peeling, with $\zeta \ne 0$, and the corresponding solution is exactly described by its WKB approximation.
The presence of modes that peel off exponentially from a causal boundary is a smoking gun of particle production. 
This happens in relativistic cases, giving rise to Hawking radiation \cite{hawking_1975}, as well as in Lorentz-violating scenarios, with an analogous phenomenon \cite{del_porro_hawking_2023}. 
Mathematically speaking, the logarithmic nonanalyticity that appears in the phase of the mode is usually associated to a thermal character of the Bogolyubov coefficients when the analytical continuation through the horizon is performed \cite{Jacobson:2003vx}.

Equivalently to the Bogolyubov coefficients approach, particle production by black holes can be described as a quantum tunnelling process \cite{Parikh_2000_HRtunneling}. 
Within a point-particle interpretation of \cref{eq:mode_omega_m}, whereby the phase $S$ plays the role of the action, one can consider the trajectory given by \cref{eq:tdot_final}, as well as its symmetric counterpart that peels in from the inside of the horizon and still solves the system \cref{eq:DRpf,eq:Epf,eq:Lpf}.
These two classically allowed paths can be connected through a quantum tunnelling process.
This boils down to considering a complex path across the horizon in order to avoid the simple pole given in \cref{eq:tdot_final}. 
In doing so, the point-particle action $S$ acquires an imaginary part that defines the tunnelling rate $\Gamma$ as
\be \label{eq:tunnelling_rate}
\Gamma=e^{ - 2 {\rm Im}( S) } \,,
\ee
where ${\rm Im}( S)$ indicates the imaginary part of the action. 

This rate, if the geometry is asymptotically flat and the matter quantum state is taken to be vacuum at the horizon for the infalling observer, provides information on the particle distribution on $\mathscr{I}^+$. 
In particular, if $\Gamma$ is of the form
\be \label{eq:boltzmann_Gamma}
\Gamma=e^{ -\Omega/T } \,,
\ee
with a $T$ that does not dependent on $\Omega$, then one can immediately infer a thermal distribution with a temperature $T$ \cite{DelPorro_2024_tesi}. 
Of course, the actual spectrum detected at infinity will also be affected by a greybody factor, namely by the probability of a particle, after being produced, to climb the gravitational potential up to the asymptotic region \cite{Jacobson:2003vx}. 
However, this greybody factor does not spoil the thermal character of the distribution. 

The tunnelling method has the great advantage of relying only on a local analysis around the horizon, which renders this formalism particularly well suited for the present case. 
As we have seen, the $\zeta=0$ trajectories maximally violate the WKB approximation, so they will not contribute to the tunnelling amplitude. 
Therefore, in what follows, we will consider only the case in which $\zeta \ne 0$ and $\kappa$ is finite. 

Along the same line of thought of \cite{del_porro_gravitational_2022,del_porro_hawking_2023} we consider the trajectory \eqref{eq:tdot_final}  assigning a small imaginary part $+ i \varepsilon$ to $\bar \tau$ in order to shift the pole from the real axis:
\be \label{eq:dt_imaginary}
\rmd t = \frac{1}{\kappa} \frac{\rmd \tau}{\tau - \bar \tau - i \varepsilon} \,.
\ee
Therefore, the point-particle action becomes:
\be \label{eq:S_nearS}
S=- \Omega \int \left[ \rmd t - \frac{1}{\kappa} \frac{\rmd \tau}{\tau - \bar \tau - i \varepsilon} \right]\,,
\ee
which is correctly constant when evaluated on \cref{eq:dt_imaginary}. 
The imaginary part of $S$ can be computed just by integrating on a path which crosses $\mathcal S$ and then considering the limit $\varepsilon \to 0^+$. 
Any regular path $t(\tau)$ connecting $\tau_1 < \bar \tau$ with $\tau_2 > \bar \tau$ gives the same result. 
So, without any loss of generality, we can take the an \ae{}ther integral line, for which, near $\mathcal S$, $\rmd t=0$. 
Thus, we have
\be \label{eq:ImS}
{\rm Im}(S)= \lim_{\varepsilon \to 0^+} {\rm Im} \left[ \frac{\Omega  }{\kappa} \int_{\tau_1}^{\tau_2}  \frac{\rmd \tau}{\tau - \bar \tau - i \varepsilon} \right]= \frac{\pi \Omega}{\kappa} \,,
\ee
from which we can read off the tunnelling rate. 
Such rate turns out be of the form \eqref{eq:boltzmann_Gamma} with temperature
\be \label{eq:T_S}
T_{\mathcal{S}}= \frac{\kappa}{2 \pi}= \frac{1+\zeta^2}{ \zeta^2 } \frac{(\mathfrak{a} \cdot \chi)}{2 \pi} \biggr|_{\mathcal{S}} \,.
\ee

Since $T_\mathcal{S}$ is proportional, up to a numerical factor $1/(2 \pi)$, to the peeling factor $\kappa$, the same observations made in section \ref{s:kappa_peeling} hold at this level. 
The limit $\zeta^2 \to \infty$, which recovers the spherically symmetric result, works here as a consistency check. 
In general, $T_{\mathcal{S}}$ is nonconstant on the horizon: this seems to point towards an out-of-thermal-equilibrium surface. 

Remarkably, if we require the twist $\omega_{ab}$ to vanish quadratically on $\mathcal{S}$ (thus making the system \qu{slightly more} hypersurface orthogonal), then $(\mathfrak{a} \cdot \chi)$ is constant on the horizon. 
This seems a step forward in the direction of a formulation of a \emph{zeroth law} of thermodynamics for universal horizons without hypersurface orthogonality.  

However, even in that case, the $\zeta$-dependence remains. 
This seems to describe --- at fixed energy $\Omega$ --- an enhancement of the production along the $\htheta$-direction. 
In other words, if $\zeta$ is kept free, a particle with Killing energy $\Omega$ is produced by the black hole at some $\zeta^2 < \infty$ with the same probability as a particle with energy $\Omega'$ is produced at $(\zeta')^2=\infty$ if
\be
\Omega= \frac{\zeta^2}{1+\zeta^2} \Omega' \,.
\ee 
As explained in section \ref{s:fixing_zeta}, a possible way to get around the problem is to link $\zeta$ to purely geometrical quantities, such as the acceleration $\mathfrak{a}_a$ and its derivative. 
That would translate the thermal equilibrium of $\mathcal S$ into a geometrical requirement on the background. 
From \eqref{eq:zeta_bound}, a sufficient condition to ensure a constant $T_{\mathcal S}$ is\footnote{This would set $\zeta_1^2=\zeta_2^2=\infty$.}
\be \label{eq:conditions_0_law}
\eval{(\mathfrak a \cdot \htheta)}_{\mathcal{S}}=0 
\qq{and}
\eval{u^a \partial_a( \htheta^b \omega_{bc} \chi^c)}_{\mathcal{S}}=0 \,.
\ee
This tells us, once again, that in order to really probe the thermodynamics of universal horizons, the system needs some dynamical input. 

Let us emphasise that \cref{eq:conditions_0_law} is also informative for the hypersurface orthogonal case. 
If we take $u_a$ to be everywhere twist free, the condition $\omega_{ab}=0$ becomes redundant in the definition \eqref{eq:defUH} of the universal horizon, and so does the condition on the right of \cref{eq:conditions_0_law}. 
A consistent thermodynamics, however, would still need the axisymmetric hypersurface orthogonal solution to satisfy $(\mathfrak a \cdot \htheta)|_{\mathcal{S}}=0$.

Whether this dynamical requirement could be fulfilled is still an open question: here our analysis has focussed only on the kinematics, while a full characterisation of the thermodynamics of rotating universal horizons is left to future works.

\section{Conclusions}

In this work, we have explored the existence and properties of universal horizons in theories where Lorentz invariance is broken by a preferred vector field that is not necessarily hypersurface orthogonal. 
While previous studies have largely focused on cases where hypersurface orthogonality holds, we have provided a generalised characterisation of universal horizons that extends beyond this constraint. 
Our approach was motivated by the recently discovered stealth Kerr solution in Einstein--\ae{t}her gravity, which features a quasi universal horizon --- a surface satisfying key conditions of a universal horizon, but failing to be truly so due to the presence of a nonvanishing \ae{t}her twist.  

Through an analysis of causality for Lorentz-violating degrees of freedom, we derived necessary conditions for a surface to act as a causal boundary for all modes, regardless of their propagation speed. 
We demonstrated that, under certain conditions, a universal horizon can still emerge even in the absence of hypersurface orthogonality, provided the twist vanishes locally at the horizon. 

We have then revisited in the light of these results the quasi universal horizon of the recently found stealth Kerr solution, and showed how it can be adjusted to form a proper universal horizon by a simple local deformation of the \ae{t}her field.

We further investigated the radiative properties of these universal horizons, by analysing the behaviour of singular trajectories usually associated to Hawking radiation. 
After identifying which of such trajectories are kinematically and causally allowed, we have shown their role in determining the Hawking temperature. 
We found that generic universal horizons radiate with a temperature that depends on the position on the horizon from which the radiation originates and on some properties of the \ae{t}her.

The emission seems to be highly enhanced in the angular direction, which could be in conflict with the existence of a zeroth law for these black holes and might even be the source of instabilities. 
Nonetheless, we have found under which conditions such temperature would be indeed constant on the whole horizon. 
This, in turn, implies that the analysis presented here might also be used as a guide towards the definition of a healthy thermodynamics for axisymmetric universal horizons.

In conclusion, we hope that these results, beyond their intrinsic relevance to the field, will stimulate further explorations. 
A key avenue would be to study the global structure of solutions featuring universal horizons beyond hypersurface-orthogonal setups, including their stability and observational signatures. 
Additionally, studying the interplay between universal horizons and quantum field theory effects in curved spacetime might provide deeper insights into the semiclassical behaviour of black holes in Lorentz-violating gravity and the resilience of black hole thermodynamics in this setting.  
We plan to dwell further into these themes in future investigations.


\acknowledgments

We want to thank Marc Schneider and Eugeny Babichev for useful discussion. 
FDP acknowledges support of the research grant (VIL60819) from VILLUM FONDEN.
JM acknowledges support of ANR grant StronG (ANR-22-CE31-0015-01).


\appendix

\section{Hamiltonian analysis of particles with nonlinear dispersion relations}
\label{a:hamiltonian}

The goal of this appendix is to recap some results on the Hamiltonian dynamics of particles satisfying a modified dispersion relation of the kind treated in this paper.
In particular, we wish to show that the trajectory followed by such particles is not a geodesic of the curved background spacetime, but rather a modification thereof.
We then explain how, despite this modification, Killing symmetries can still lead to constants of motion.
The analysis is carried out along the lines of e.g.~\cite{BarcaroliHamiltonGeometry2015,FrolovBlackHoles2017}.

We consider a point particle's phase space charted by the coordinates $(X^a, k_a)$, where $X^a$ represents the particle's position in spacetime while $k_a$ is the momentum conjugate to $X^a$.
In accordance with the discussion of \cref{s:causality}, we assume that the particle's motion is governed by the Hamiltonian
\be
H = \frac{1}{2} \left[g^{ab} k_a k_b + \Lambda^2 \mathcal{G} \left( \abs{\vb{k}}^2 /\Lambda^2 \right)\right]\, ,
\ee
where $\mathcal{G}$ is some function and recall that $\abs{\vb{k}}^2 = P^{ab} k_a k_b$; 
this writing is thus very general, and \cref{eq:DR} corresponds to the particular choice
\be
\mathcal{G} = \sum_{j=2}^N \frac{\beta_j}{\Lambda^{2j}} \left( P^{a b} k_a k_b \right)^j \, .
\ee

For any two phase space functions $A$ and $B$, their Poisson bracket is defined as customary,
\be
\pb{A}{B} := \pdv{A}{X^a} \pdv{B}{k_a} - \pdv{B}{X^a} \pdv{A}{k_a}\, ,
\ee
and the derivative of any such function along the particle's trajectory is 
\be
\dot{A} = \pb{A}{H}\, .
\ee
Hence, in particular, the value of the Hamiltonian is conserved on shell: 
for consistency with the discussion of \cref{s:causality}, we assume that the Hamiltonian vanishes --- i.e.~that the particle is massless ---, but the extension is straightforward.

Under this assumption, we can compute the phase and the group velocity in the preferred frame, to get
\be
c_{\rm p} &= \sqrt{1 + \Lambda^2 \mathcal{G}/\abs{\vb{k}}^2} \, ,\\
c_{\rm g} &= \frac{1}{c_{\rm p} } \left[1 + \mathcal{G}' \right]\, ,
\ee
where $\mathcal{G}'$ is the derivative of $\mathcal{G}$ with respect to its argument.
As one might expect, these velocities appear in Hamilton's equations, which read (cf.~\cref{eq:Xdot})
\be
\label{eq:Hamilton_X}
\Dot{X}^a &= g^{ab} k_b + \left( c_{\rm g} c_{\rm p} - 1 \right) P^{\mu \nu} k_b \nonumber\\
&= c_{\rm g} c_{\rm p} k^a - \omega \left( c_{\rm g} c_{\rm p} - 1 \right) u^a 
\, ,\\
\label{eq:Hamilton_k}
\Dot{k}_a &= - \frac{1}{2} k_b k_c \left[ \partial_a g^{bc} + \left( c_{\rm g} c_{\rm p} - 1 \right) \partial_a P^{bc}\right] \nonumber\\
&= - \frac{1}{2} k_b k_c \left[ c_{\rm g}c_{\rm p} \partial_a g^{bc} + \left( c_{\rm g} c_{\rm p} - 1 \right)  \partial_a \left( u^b u^c \right) \right]
\, .
\ee
We remark that, as mentioned in the main text (cf.~\cref{eq:ell}), $k^a$ is not proportional to the velocity $\dot{X}^a$, i.e.~it is not tangent to the trajectory.

Noting that $\dot{k}_a = \Dot{X}^b \partial_b k_a$, the two Hamilton's equations can be compounded into
\be
\left[ k^b \partial_b k_a + \frac{1}{2} k_b k_c \partial_a g^{b c} \right] = - \left( c_{\rm g} c_{\rm p} - 1 \right) \left[ k_b P^{b c} \partial_c
k_a + \frac{1}{2} k_b k_c \partial_a P^{b c} \right] \, .
\ee
The first square bracket can be repackaged into a covariant derivative using standard methods, so this equation takes the form of a \qu{modified} geodesic equation 
\be\label{eq:geodesic_mod_1}
k^b \nabla_b k_a = A_a\, ,
\ee
where
\be
A_a :=  \left( 1 -  c_{\rm g} c_{\rm p}\right) \left[ k_b P^{b c} \partial_c
k_a + \frac{1}{2} k_b k_c \partial_a P^{b c} \right]
\ee
can be interpreted as an acceleration.
At least formally, it might be possible to write $A_a \propto k_b P^{b c} \tilde{\nabla}_c k_a$, where $\tilde{\nabla}_a$ is a \qu{covariant derivative} constructed out of the projector.
However, we find it more illuminating to manipulate \cref{eq:geodesic_mod_1} into
\be
k^b \nabla_b k_a &= \omega \left( 1 - \frac{1}{c_{\rm g}c_{\rm p}} \right) \left[ u^b \partial_b k_a + k_b \partial_a u^b \right] \nonumber\\
\label{eq:geodesic_mod_2}
&= \omega \left( 1 - \frac{1}{c_{\rm g}c_{\rm p}} \right) \Lie{u}{k_a}\, .
\ee
Clearly, this acceleration vanishes when $c_{\rm g} c_{\rm p} = 1$.
Formally, it also vanishes for trajectories whose momenta satisfy $\omega = - (u \cdot k) = 0$; according to \cref{eq:Hamilton_X}, these trajectories are themselves orthogonal to the \ae{}ther, but they are therefore not causal according to the definition of \cref{s:causality}.
Further note that $A_a$ is not orthogonal to $k_a$, since the norm of this vector, on its own, is not conserved along the motion.

These remarks motivate us to investigate constants of motion on more general grounds.
In analogy with \cite{FrolovBlackHoles2017}, we consider a symmetric tensor field $K^{a_1 \dots a_n}$ that depends on the position $X^a$ but not on the momentum $k_a$.
The quantity
\be
K := K^{a_1 \dots a_n} k_{a_1} \dots k_{a_n} 
\ee
is a constant of motion if its Poisson bracket with the Hamiltonian vanishes.
Computing said bracket, we find
\be
\pb{K}{H} &= k_a k_{a_1} \dots k_{a_n} \bigg\{ \left[g^{a b} \partial_b K^{a_1 \dots a_n} - \frac{n}{2} K^{b a_2 \dots a_n} \partial_b g^{a a_1} \right] \nonumber\\
&\phantom{=} + \left( c_{\rm p}c_{\rm g} - 1 \right) \left[P^{ab} \partial_b K^{a_1 \dots a_n} - \frac{n}{2} K^{b a_2 \dots a_n} \partial_b P^{a a_1} \right] \bigg\} \, .
\ee

Using similar manipulations as for the modified geodesic equation, we realise that the first square bracket can be written as the covariant derivative $\nabla^a K^{a_1 \dots a_n}$, up to terms that are antisymmetric in their indices and thus vanish when contracted with the momenta; this term gives rise to the usual Killing equation.
As for the second square bracket, it might still be possible to write it in terms of a new \qu{covariant derivative}, but as before we find it more useful to massage the equation above into
\be
\pb{K}{H} &= k_a k_{a_1} \dots k_{a_n} \bigg\{ c_{\rm p} c_{\rm g} \nabla^a K^{a_1 \dots a_n} \nonumber \\
&\phantom{=} + (c_{\rm p} c_{\rm g}-1) \left[ u^a u^b \partial_b K^{a_1 \dots a_n} - \frac{n}{2} K^{b a_2 \dots a_n} \partial_b ( u^a u^{a_1}) \right] \bigg\} \, .
\ee
Writing this last square bracket as
\be
u^a u^b \partial_b K^{a_1 \dots a_n} - \frac{n}{2} K^{b a_2 \dots a_n} \partial_b ( u^a u^{a_1}) &= u^a \left[u^b \partial_b K^{a_1 \dots a_n} - n K^{b a_2 \dots a_n} \partial_b u^{a_1} \right] \nonumber\\
&\phantom{=} + n K^{b a_2 \dots a_n} u^{[a}\partial_b u^{a_1]} \, ,
\ee
we notice that the term in the second line will vanish when contracted with the momenta;
moreover, we realise that the first term will give rise, once again, to a Lie derivative.

Hence, putting everything together, we find the following neat expression
\be
\pb{K}{H} &= k_a k_{a_1} \dots k_{a_n} \bigg\{ c_{\rm p} c_{\rm g} \nabla^a K^{a_1 \dots a_n} \nonumber\\
&\phantom{=} + (c_{\rm p} c_{\rm g}-1) u^a \Lie{u}{K^{a_1 \dots a_n}} \bigg\} \, ,
\ee
from which we deduce that if the tensor $K^{a_1 \dots a_n}$ is Killing and if it is Lie-dragged by the \ae{}ther, then it generates a constant of motion.
Notice, however, that in principle one could have other, more complicated solutions of $\pb{K}{H}=0$ in which the tensor is neither Killing nor Lie-dragged.
Coherently with the modified geodesic \cref{eq:geodesic_mod_2}, the Killing condition alone is sufficient for trajectories whose momentum is orthogonal to $u_a$.

It becomes simple to check that the rank-two Killing tensor of the Kerr spacetime is not Lie-dragged by the \ae{}ther of the stealth solution of \cite{franzin_kerr_2023}.
Hence, as anticipated, that solution does not seem to have any additional constant of motion and the motion is therefore not completely integrable.
Understanding whether one could find a different \ae{}ther flow which does Lie-drag Kerr's Killing tensor, then, becomes an intriguing question.

Moreover, it is worth pointing out that this analysis does not exclude the possibility of constants of motion that are not polynomial in the momenta.
To the best of our knowledge, this is an entirely open question already in general relativity and/or with standard dispersion relations.

\section{Raychaudhuri equations for a nongeodesic congruence}
\label{a:Raychaudhuri}

In this appendix we shall collect the Raychaudhuri equations for a timelike, nongeodesic congruence. 
In the context of relativistic hydrodynamics, these are well-known results \cite{Rezzolla_hydro}. 
Here, we want to sketch a derivation of these equations and state some identities which hold in our axisymmetric geometry. 
Let us start defining the tensor:
\be
B_{ab}= \nabla_{a}u_b \,.
\ee
As mentioned in section \ref{s:Raychaudhuri}, this tensor can be decomposed in the so-called hydrodynamical invariants
\be \label{eq:Bab_dec_app}
\nabla_a u_b= \frac{1}{3} P_{ab} \vartheta + \sigma_{ab} + \omega_{ab} - u_a \mathfrak{a}_b \,,
\ee
where the expansion $\vartheta$, the shear $\sigma_{ab}$ and the twist $\omega_{ab}$ are defined as \cite{Rezzolla_hydro}
\be 
& \vartheta= \nabla_a u^a \,,\\
& \sigma_{ab}= \nabla_{(a}u_{b)}+u_{(a} \mathfrak{a}_{b)} - \frac{1}{3} \vartheta P_{ab} \,, \\
& \omega_{ab} = \nabla_{[a}u_{b]}+u_{[a} \mathfrak{a}_{b]} \,.
\ee

The Raychaudhuri equations are nothing but the evolution of the quantities defined above, along the flow determined by $u^a$. In other words, we have to compute
\be
\dot B_{ab}= u^c \nabla_c B_{ab} \,,
\ee
where, from now on, we shall always denote the derivative along $u^a$ with a dot on top of the tensor. After some simple algebra, and making use of \eqref{eq:Bab_dec_app}
\be
\dot B_{ab}& = \nabla_a \mathfrak{a}_b - B_a^{\ c}B_{cb}+ u^c R_{cabd} u^d \nonumber \\
&= \nabla_a \mathfrak{a}_b + u_a \mathfrak{a}^c\left[ \frac{\vartheta}{3} P_{cb} + \sigma_{cb} + \omega_{cb}  \right] 
- \frac{\vartheta^2}{9}P_{ab} 
- \frac{2\vartheta}{3} \left( \sigma_{ab} + \omega_{ab} \right) \nonumber\\
&\phantom{=} - \sigma_a^{\ c} \sigma_{cb} - \omega_a^{\ c} \omega_{cb}
- \sigma_a^{\ c} \omega_{cb} - \omega_a^{\ c} \sigma_{cb} + u^c R_{cabd} u^d\,  .
\ee
Recalling that, by definition, we have
\be
& \dot{\vartheta} = g^{ab}\dot{B}_{ab}\, , \label{eq:thetadot} \\
& \dot{\sigma}_{ab} = \mathfrak{a}_a \mathfrak{a}_b + u_{(a} \dot{\mathfrak{a}}_{b)} + \dot{B}_{(ab)} - \frac{\dot{\vartheta} P_{ab} + \vartheta \dot{P}_{ab} }{3}\, , \label{eq:sigmadot} \\
& \dot{\omega}_{ab} = u_{[a} \dot{\mathfrak{a}}_{b]} + \dot{B}_{[ab]} \, , \label{eq:omegadot}
\ee
one can derive the evolution equations for a nongeodesic timelike congruence. 
In particular, the expansion
\be \label{eq:Ray_theta}
\dot{\vartheta} = \nabla_a \mathfrak{a}^a - \frac{\vartheta^2}{3} - \sigma_{ab}\sigma^{ab} +  \omega_{ab}\omega^{ab} + R_{ab}u^a u^b \,,
\ee
the shear
\be \label{eq:Ray_sigma}
\dot{\sigma}_{ab} &= - \frac{2\vartheta}{3}\sigma_{ab} - \sigma_a^{\ c} \sigma_{cb} + \omega_a^{\ c} \omega_{bc} + \frac{1}{3}P_{ab} \left[\sigma_{cd}\sigma^{cd} - \omega_{cd}\omega^{cd} \right] + u^c R_{cabd} u^d - \frac{1}{3}P_{ab} R_{cd}u^c u^d \nonumber\\
&\phantom{=} + P_{(a}^{\ c} \nabla_c \mathfrak{a}_{b)}  + \mathfrak{a}_a\mathfrak{a}_b  - \frac{1}{3}P_{ab} \left( \nabla_c \mathfrak{a}^c \right)
- \frac{\vartheta}{3} u_{(a} \mathfrak{a}_{b)} + u_{(a} \sigma_{b)c} \mathfrak{a}^c - u_{(a} \omega_{b)c} \mathfrak{a}^c \,,
\ee
and the twist
\be \label{eq:Ray_omega}
\dot{\omega}_{ab} = P_{[a}^{\ c}\nabla_c \mathfrak{a}_{b]} 
+ \frac{\vartheta}{3} u_{[a} \mathfrak{a}_{b]}+ u_{[a} \sigma_{b]c} \mathfrak{a}^c - u_{[a} \omega_{b]c} \mathfrak{a}^c  - \frac{2\vartheta}{3}\omega_{ab} - \sigma_a^{\ c} \omega_{cb} - \omega_a^{\ c} \sigma_{cb}  \,.
\ee

We stress here that it is not possible to derive a similar equation for $\mathfrak{\dot a}_a$. 
As already mentioned in section \ref{s:Raychaudhuri}, this is conceptually consistent with the fact that not every timelike congruence on a manifold defines the same flow, namely that one can not uniquely determine the evolution just by kinematical considerations. 
At the technical level, trying to apply to $\mathfrak{a}_a$ the same procedure with which we derived \cref{eq:Ray_theta,eq:Ray_sigma,eq:Ray_omega}, gives back a mathematical identity without any additional information on the evolution of the hydrodynamical invariants.

\subsection{Axisymmetric spacetimes: a useful identity}

If the spacetime is axisymmetric (in the sense of \cref{eq:uLie}), the twist tensor $\omega_{ab}$ has a single independent component $(\hat \theta^a \omega_{ab} \chi^b)$. 
Its evolution is given by
\be
u^c \nabla_c \left( \hat \theta^a \omega_{ab} \chi^b \right) = \omega_{ab} u^c \nabla_c \left( \hat \theta^a \chi^b \right) + \hat \theta^a \chi^b \dot \omega_{ab} \,.
\ee
Using \cref{eq:Ray_omega} and the orthogonality between $\hat \theta^a$ and $u^a$, one can simply rewrite:
\be
\hat \theta^a \chi^b \dot \omega_{ab} & = \frac{1}{2} \hat \theta^c \nabla_c \left( \mathfrak{a} \cdot \chi \right)- \frac{1}{2} \hat \theta^c \mathcal{L}_\chi \mathfrak{a}_c- \frac{u \cdot \chi}{2} \left[ \hat \theta^b u^c \nabla_c \mathfrak{a}_b+ \left(\sigma_{ac} + \omega_{ac} \right) \hat{\theta}^a \mathfrak{a}^c \right] \nonumber\\
&\phantom{=} + \omega_{cd} \left[ - \frac{2 \vartheta}{3} \chi^d \hat \theta^c - \htheta^a \sigma_a^{\ c} \chi^d - \htheta^c \sigma^d_{\ b}\chi^b \right] \,.
\ee
Let us notice that, if $\chi$ is a Killing vector, which Lie-drags $u_a$, we have $\theta^c \mathcal{L}_\chi \mathfrak{a}_c=0$. 
The interesting part of the equation above can be obtained in the presence of a universal horizon. 
If \cref{eq:defUH} happens to be satisfied on some surface $\mathcal S$, we can compute
\be \label{eq:omega_dot_UH}
\eval{u^c \nabla_c \left( \hat \theta^a \omega_{ab} \chi^b \right)}_{\mathcal S} = \eval{\frac{1}{2} \hat \theta^c \nabla_c \left( \mathfrak{a} \cdot \chi \right)}_{\mathcal S} \,.
\ee

Equation \eqref{eq:omega_dot_UH} has a geometrical meaning: since $(\mathfrak{a} \cdot \chi)$ represents the normal derivative of $(u \cdot \chi)$, the vanishing of its $\hat \theta$-derivative means that the normal vector $n_a= \nabla_a (u \cdot \chi)$ is constant along $\mathcal S$. 
This is of particular relevance for section \ref{s:kappa_peeling}, where one can directly see that $(\mathfrak{a} \cdot \chi)$ enters in the definition of the peeling surface gravity $\kappa$.

\section{Calculations on the stealth Kerr solution}
\label{a:stealthKerr}

In this appendix we report the computations that we omitted in \cref{s:stealthKerr}.
Namely, we will prove that, in the stealth Kerr solution, the surface $r=r_\quh$ is either spacelike or, in a limiting case, null.
As mentioned in the main text, this boils down to establishing that the following inequality holds:
\be
2 \abs{\frac{\htheta^\mu \, \omega_{\mu \nu} \, s^\nu}{ \left( \mathfrak{a} \cdot s \right)} }_{r=r_\quh } = 
2 \abs{\frac{\htheta^\mu \, \omega_{\mu \nu} \, \chi^\nu}{ \left( \mathfrak{a} \cdot \chi \right)} }_{r=r_\quh } < 1 \, .
\ee

In the stealth Kerr solution, we have
\be
\htheta^\mu \, \omega_{\mu \nu} \, \chi^\nu &= \frac{\partial_\theta u_t}{2\sqrt{g_{\theta \theta}}} 
\qq{and} 
\mathfrak{a}_\rho\, \chi^\rho = u^r \partial_r u_t\, .
\ee
Moreover, from the condition $( u \cdot u) = -1$ we deduce $\eval{u^r}_{r=r_\quh } = \eval{ (\sqrt{-g_{rr}} )^{-1/2} }_{r=r_\quh }$; 
so, exploiting the properties of the (inverse) Kerr metric, the inequality above becomes
\be\label{eq:condKerr1}
\abs{\frac{\sqrt{g^{\theta \theta}} \partial_\theta u_t}{ \sqrt{\abs{g^{rr}}}\partial_r u_t} }_{r=r_\quh } < 1\, .
\ee

The previous expression contains an indeterminate form of the type $[0]/[0]$.
To see this, note that calling $F:= \Delta \Sigma - \Delta_\quh \Sigma_\quh$ gives (cf.~\cref{eq:uStealth})
\be
\partial_\mu u_t = \mp \frac{1}{2} \sqrt{\frac{A}{F}} \left[ \frac{\partial_\mu F}{A} - F \frac{\partial_\mu A}{A^2} \right]\, .
\ee
On the (candidate) quasi universal horizon, we have $\eval{F}_\quh = \eval{\partial_\mu F}_\quh = 0$ and $\eval{A}_\quh \neq 0$;
hence the derivative reduces to
\be
\eval{\partial_\mu u_t}_{r=r_\quh } = \eval{ \mp \frac{1}{2} \frac{\partial_\mu F}{\sqrt{A F}} }_{r=r_\quh }\, ,
\ee
which is already indeterminate.
To verify \cref{eq:condKerr1}, we need to evaluate the ratio $\partial_\theta u_t/ \partial_r u_t$, which --- one might guess --- will ultimately be given by $\partial_\theta F/\partial_r F$.
However, this is still an indeterminate form.
Hence, what one would need to do is expand $F$ and its first derivatives to second order in a Taylor expansion around a point on the quasi universal horizon, then compute the ratio, and finally take a limit.

Such tedious, albeit somewhat instructive computation can be avoided by noting that $u_t \left( r_\quh(\theta), \theta \right)$ is a constant function of $\theta$.
Hence it must be that
\be
0 &= \dv{}{\theta} \left[ u_t\left( r_\quh(\theta), \theta \right) \right] \nonumber \\
&= \left[ \dv{r_\quh}{\theta} \partial_r u_t + \partial_\theta u_t \right]_{r=r_\quh(\theta)}
\ee
and therefore
\be
\eval{ \frac{\partial_\theta u_t}{\partial_r u_t} }_{r=r_\quh(\theta)} = - \dv{r_\quh}{\theta}\, .
\ee

Clearly, one would obtain the same result by employing the \qu{brute force} method outlined above, only to realise that the Hessian matrix of $F$, evaluated on $r=r_\quh(\theta)$, has a vanishing eigenvalue --- corresponding to an eigenvector that is tangent to the quasi universal horizon.
The \qu{smart} method instead, besides being neater and faster, exemplifies the emergence of nontrivial identities, involving $\dv*{r_\quh}{\theta}$, that arise by taking total derivatives with respect to $\theta$; 
in practice, it is thanks to these identities, and the simplifications they allow, that one is able to perform computations without ever needing the explicit expression of $r_\quh (\theta)$.

One of such identities gives
\be
\dv{r_\quh}{\theta} = \sin(2\theta) \frac{a^2 \left( r_\quh - M \right)}{ 6 r_\quh \left( r_\quh - M \right) + a^2\left(1+\cos^2\theta \right)}\, ,
\ee
which is a rewriting of \cite[eq.~(32)]{franzin_kerr_2023} and derives from differentiating \cref{eq:rquh}.
Using this expression, along with the explicit form of the metric components, we arrive at the final version of the inequality we must prove:
\be\label{eq:condKerr}
\abs{\sin(2\theta) \frac{r_\quh-M}{\sqrt{\abs{\Delta_\quh} }} \frac{a^2}{6 r_\quh \left( r_\quh - M \right) + a^2\left(1+\cos^2\theta \right)} } < 1 \, .
\ee

One can prove, thanks again to \cref{eq:rquh} (cf.~\cite{franzin_kerr_2023}), that $r_\quh \geq M$, so the last term is manifestly smaller than one; 
moreover, using once more \cref{eq:rquh}, one can write
\be
\frac{r_\quh-M}{\sqrt{ \abs{\Delta_\quh} }} = \sqrt{\frac{r_\quh^2 -r_\quh M }{\Sigma_\quh}} < 1\, ;
\ee
finally, the sine is obviously not grater than one in absolute value.
Hence each term of the left-hand side, on its own, satisfies or at most saturates the bound and therefore the inequality is satisfied.

More precisely, $r=r_\quh$ is a spacelike surface whenever $a \leq M$.
The limit $a \to M$, instead, should be handled with some care:
the bound is still satisfied, but one finds
\be
\eval{ \left(\mathfrak{a} \cdot \chi\right) }_{r = r_\quh} \to 0
\qq{and therefore also}
\eval{\htheta^\mu \, \omega_{\mu \nu} \, \chi^\nu}_{r = r_\quh} \to 0
\, ;
\ee
confronting with \cref{eq:normn}, this entails that the surface $r=r_\quh$ becomes null.
Indeed, in this limit one has that $r_\quh \to M$, which is also the location of the (degenerate) Killing horizon.
Hence, in this limit $r=r_\quh$ is no longer a quasi universal horizon but rather an \qu{extremal} universal horizon as per~\cite{FranchiniBlackHole2017}.



\bibliographystyle{JHEP}
\bibliography{biblio.bib}

\end{document}